\documentclass{article}
\usepackage{natbib}
\usepackage[final]{neurips_data_2022}
\usepackage[utf8]{inputenc} 
\usepackage[T1]{fontenc}    
\usepackage{booktabs}       
\usepackage{amsfonts}       
\usepackage{nicefrac}       
\usepackage{microtype}      
\usepackage{graphicx}
\usepackage{booktabs}
\usepackage{multirow}

\usepackage{pifont}
\newcommand{\cmark}{\ding{51}}%
\newcommand{\xmark}{\ding{55}}%
\usepackage{subcaption} 
\usepackage{wrapfig}
\usepackage{algpseudocode}
\usepackage{etoolbox}
\BeforeBeginEnvironment{wrapfigure}{\setlength{\intextsep}{4pt}}
\usepackage{textcomp}
\usepackage{xcolor}         
\usepackage{hyperref}       
\hypersetup{colorlinks,allcolors=black}
\usepackage{url}            

\title{OADAT: Experimental and Synthetic Clinical Optoacoustic Data for Standardized Image Processing}

\author{%
\centerline{Firat Ozdemir$^{1}$%
\thanks{Equal contribution. Author ordering determined by coin flip. %
For correspondence, please contact either firat.ozdemir@datascience.ch or berkan.lafci@uzh.ch.}%
\quad \quad Berkan Lafci$^{2,3*}$ \quad \quad Xos\'e Lu\'is De\'an-Ben$^{2,3}$} 
\vspace{0.2em}
\\
\centerline{\textbf{Daniel Razansky}$^{2,3}$ \quad \quad \textbf{Fernando Perez-Cruz}$^{1,4}$} 
\vspace{0.4em}
\\
$^1$Swiss Data Science Center, ETH Zurich and EPFL, Zurich, Switzerland\\
$^2$Institute of Pharmacology and Toxicology and Institute for Biomedical Engineering, \\ \indent
Faculty of Medicine, University of Zurich, Switzerland \\
$^3$Institute for Biomedical Engineering, Department of Information Technology and \\ \indent
Electrical Engineering, ETH Zurich, Switzerland \\ 
$^4$Institute for Machine Learning, Department of Computer Science, ETH Zurich, Switzerland
}

\begin{document}

\maketitle

\begin{abstract}
Optoacoustic (OA) imaging is based on excitation of biological tissues with nanosecond-duration laser pulses followed by subsequent detection of ultrasound waves generated via light-absorption-mediated thermoelastic expansion.
OA imaging features a powerful combination between rich optical contrast and high resolution in deep tissues.
This enabled the exploration of a number of attractive new applications both in clinical and laboratory settings.
However, no standardized datasets generated with different types of experimental set-up and associated processing methods are available to facilitate advances in broader applications of OA in clinical settings.
This complicates an objective comparison between new and established data processing methods, often leading to qualitative results and arbitrary interpretations of the data.
In this paper, we provide both experimental and synthetic OA raw signals and reconstructed image domain datasets rendered with different experimental parameters and tomographic acquisition geometries.
We further provide trained neural networks to tackle three important challenges related to OA image processing, namely accurate reconstruction under limited view tomographic conditions, removal of spatial undersampling artifacts and anatomical segmentation for improved image reconstruction.
Specifically, we define 44 experiments corresponding to the aforementioned challenges as benchmarks to be used as a reference for the development of more advanced processing methods.
\end{abstract}

\section{Introduction}
\begin{figure}
\centering
\includegraphics[width=0.96\textwidth]{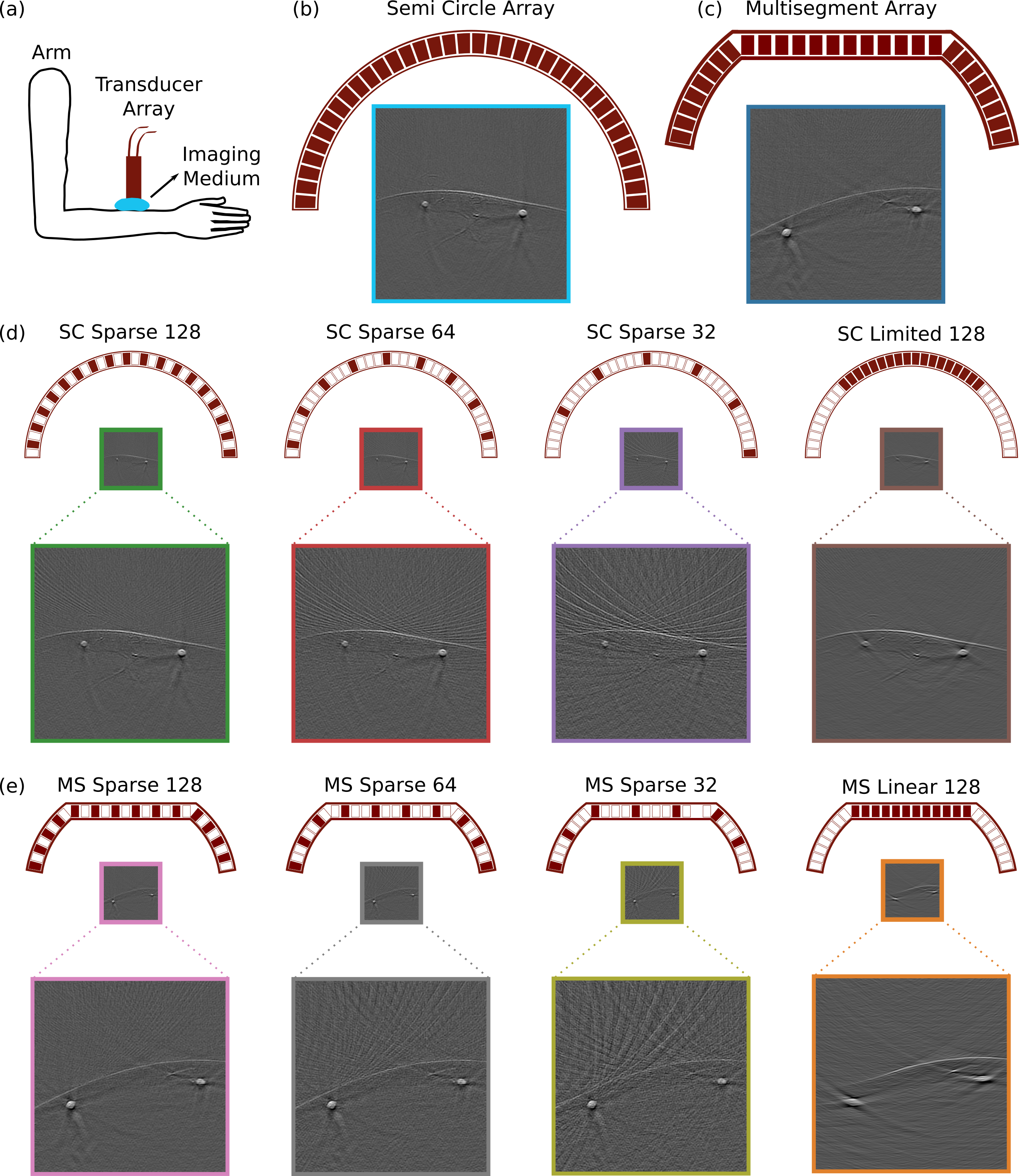}
\caption{
Experimental data acquisition, transducer arrays and resulting images.
(a) Experimental setup for optoacoustic forearm imaging.
(b) Semi circle array along with an example of acquired images.
(c) Multisegment array along with an example of acquired images.
(d) Uniform subsampling for the semi circle array (128, 64 and 32 elements) and limited view acquisition for the semi circle array with reduced angular coverage (128 elements).
(e) Uniform subsampling for the multisegment array (128, 64 and 32 elements), and linear array acquisition for the multisegment array (128 elements). 
Transducer elements are shown as actively receiving (red) or off (white).
}
\label{fig:experimental_data}
\end{figure}

Optoacoustic (OA) imaging is being established as a powerful method with increasing application areas in clinical \citep{steinberg2019photoacoustic, su2010advances} and preclinical settings~\citep{lafci2020noninvasive, mercep2015whole}.
Using nanosecond-duration pulsed lasers operating in the visible and near-infrared (NIR) optical wavelength range, biological tissues are thermoelastically excited.
This excitation yields ultrasound (US) waves, from which OA images are tomographically reconstructed (Fig.\,~\ref{fig:experimental_data}a).
The rich optical contrast from endogenous tissue chromophores such as blood, melanin, lipids and others are combined with high US resolution, i.e.,\ tens of micrometers.
This unique feature makes OA particularly suitable for molecular and functional imaging.
Other important advantages such as the feasibility of hand-held operation, the fast acquisition performance (real time feedback) and the non-invasive safe contrast (i.e., non-ionizing radiation) further foster the wide use of OA in multiple biomedical studies.
OA imaging has been shown to provide unique capabilities in studies with disease models e.g., of breast cancer~\citep{manohar2019current, diot2017multispectral, butler2018optoacoustic}, as well as for the clinical assessment of Crohn’s disease~\citep{knieling2017multispectral}, atherosclerotic carotid plaques~\citep{karlas2021multispectral} or skin cancer~\citep{deanben2021optoacoustic}.
As the range of applications of OA imaging gets broader, the need for different data processing pipelines increases in parallel.
Also, new methods are continuously being developed to provide an enhanced OA performance.
Specific examples include increased temporal resolution with compressed/sparse data acquisitions~\citep{ozbek2018optoacoustic}, accurate image reconstruction algorithms~\citep{deanben2012accurate}, light fluence correction by segmenting the tissue boundaries~\citep{lafci2021deep} or enhanced spectral unmixing algorithms from multispectral data~\citep{tzoumas2017spectral}.

Three major challenges suitable for data-driven approaches in clinical OA imaging are summarized below:
\\
\textbf{Sparse acquisition:} 
OA imaging provides a unique potential to monitor fast-changing events such as cardiac arrhythmias~\citep{ozsoy2021ultrafast}, neuronal activity~\citep{robin2021hemodynamic} or indocyanine green clearance~\citep{grunherz2022preoperative} \textit{in vivo}.
For this, ultra-fast imaging systems capable of capturing changes in living organisms occurring at up to millisecond temporal scales are required.
The main limiting factor affecting the achievable frame rate is the data transfer capacity.
This limitation can be eliminated by reducing the number of acquired channels (signals).
Therefore, sparse or compressed sensing methods have been proposed both using conventional methods~\citep{ozbek2018optoacoustic} and deep learning algorithms~\citep{davoudi2019deep}.
\\
\textbf{Limited view reconstruction:} 
OA is inherently a tomographic imaging modality.
Acquisition of pressure signals from different angles is essential to capture the information encoded in US waves traveling in a 3D medium in order to render accurate tomographic reconstructions.
This further increases the image contrast, resolution and quantitativeness.
However, tomographic coverage of the samples is often hindered by physical restrictions.
Thereby, new image processing pipelines have been suggested to improve limited-view-associated challenges in OA imaging by using data-driven algorithms in the image domain~\citep{guan2020limited}, signal domain~\citep{klimovskaia2022signal} and combination of both domains~\citep{davoudi2021deep, lan2019ynet}.
\\
\textbf{Segmentation:} 
Optimal OA reconstruction algorithms need to account for different optical and acoustic properties in biological tissues and in the coupling medium (water).
For example, the speed of sound (SoS) depends on the elastic properties of the medium.
Proper assignment of SoS values in tissues and in water requires accurate delineation of the tissue boundaries.
Thereby, segmentation of structures~\citep{lafci2021deep, schellenberg2021semantic} in OA images has been shown to enhance the image reconstruction performance.
Additionally, the optical fluence (intensity) also varies with depth across different tissues.
This issue remains as one of the main factors affecting quantification in OA images~\citep{brochu2017towards} and can also be corrected with tissue segmentation~\citep{mandal2016visual}.

As an emerging method, OA imaging requires standardization, open source code publication and data sharing practices to expedite the development of new application areas and data processing pipelines.
Particularly, the aforementioned challenges associated to high data throughput, limited angular coverage, SoS assignment and fluence corrections require coordinated efforts between experimental and data science communities.
Initial efforts to standardize data storage formats and image reconstruction algorithms have been undertaken~\citep{grohl2022ipasc}.
However,~\cite{grohl2022ipasc} focus on standard reconstruction algorithms and propose data storage formats for acquisition related metadata.
Data-driven image and signal processing methods require additional initiatives on fast access to large bulks of OA image and signal data and benchmarks for learning-based methods.

Here, we provide experimental data and simulations of forearm datasets as well as benchmark networks aiming at facilitating the development of new image processing algorithms and benchmarking.
These ``Experimental and Synthetic Clinical Optoacoustic Data (OADAT)'' include,
(i) large and varied clinical and simulated forearm datasets with paired subsampled or limited view image reconstruction counterparts,
(ii) raw signal acquisition data of each such image reconstruction,
(iii) definition of 44 experiments with gold standards focusing on the aforementioned OA challenges,
(iv) pretrained model weights of the networks used for each task, and
(v) user-friendly scripts to load and evaluate the networks on our datasets.
The presented datasets and algorithms will expedite the research in OA image processing.
\section{Background}
For OA imaging, the objects are excited with the nanosecond-duration laser pulses in visible or NIR light wavelengths which result in thermoelastic expansion of the structures.
This expansion generates pressure waves (US signals) that are detected by transducer arrays.
Corresponding images are reconstructed by solving the OA inverse problem on the acquired signals.
Below, we explain the transducer arrays used for data acquisition, the reconstruction algorithm used to generate images from acquired signals and the sampling/acquisition techniques.
Detailed explanation about OA imaging and used tools can be found in Appendix \ref{sec:optoacoustic_imaging_background}.
\subsection{Transducer arrays}
\label{sec:transducer_arrays}

\begin{figure}
\centering
\includegraphics[width=1.\textwidth]{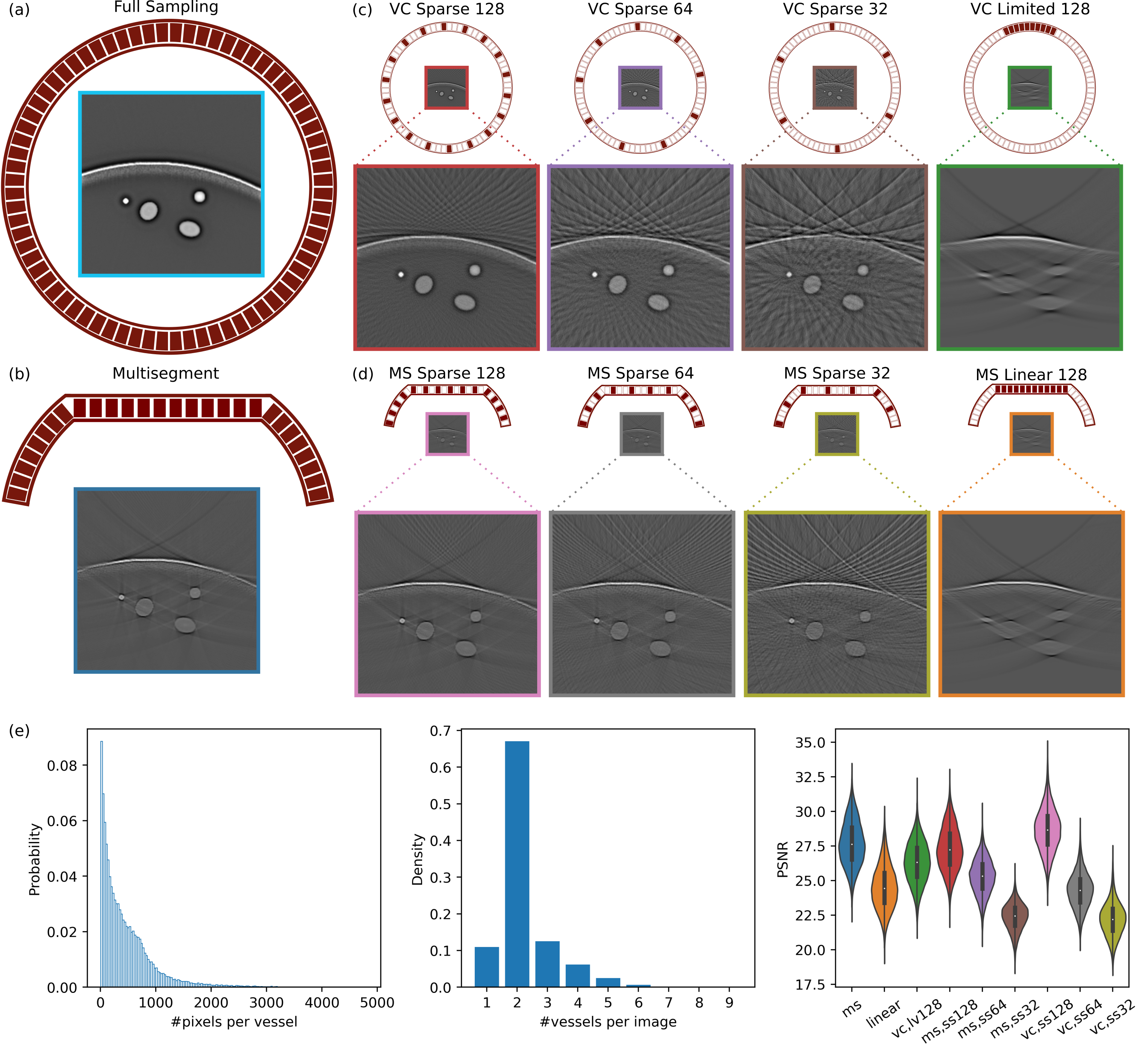}
\caption{Overview of the simulated data.
(a) Virtual circle array and an image reconstructed using 1,024 transducer elements.
(b) Multisegment array and the corresponding image reconstructed using combined linear and concave parts of the transducer array.
(c) Uniform subsampling of virtual circle array with 128, 64 and 32 elements and limited view acquisition with reduced angular coverage (128 elements).
(d) Uniform subsampling of multisegment array with 128, 64 and 32 elements and linear array acquisition (128 elements).
(e) Vessel size distribution (pixels per vessel), number of vessels per image, and peak signal-to-noise ratio of full sampling compared to other reconstructions (x axis naming conventions are explained in Sec.~\ref{sec:scd}).
Transducer elements are shown as actively receiving (red) or off (white).
}
\label{fig:simulated_data}
\end{figure}

\textbf{{Semi circle}} array contains 256 transducers elements distributed equidistantly over a semi circle (concave surface, Fig.~\ref{fig:experimental_data}b).
\textbf{{Multisegment}} array is a combination of linear array in the center and concave parts on the right and left sides, designed to increase angular coverage, as shown in Fig.~\ref{fig:experimental_data}c.
The linear part contains 128 elements and each of the concave parts consist of 64 elements, totaling to 256.
\textbf{{Linear}} array is the central part of the multisegment array with 128 elements (Fig.~\ref{fig:experimental_data}c).
The array geometry is optimized for US data acquisitions with planar waves.
Hence, it produces OA images with limited view artifacts due to reduced angular coverage which is a limiting factor for OA image acquisitions.
\textbf{{Virtual circle}} array is generated to simulate images with 360 degree angular coverage and yields artifact free reconstructions (Fig.~\ref{fig:simulated_data}a).
It contains 1,024 transducer elements distributed over a full circle with equal distance.
We also have a virtual multisegment array that correspond to its physical counterpart.
Additional geometric and technical details are listed in Appendix \ref{sec:transducer_array_details}.

\subsection{Reconstruction method}
\label{sec:reconstruction_methods}
We use backprojection algorithm in this study to generate OA images from the acquired signals\footnote{Python module for OA reconstruction: \href{https://github.com/berkanlafci/pyoat}{github.com/berkanlafci/pyoat}.}.
This algorithm is based on delay and sum beamforming approach \citep{ozbek2013realtime} (see Appendix \ref{sec:oa_image_reconstruction} for details). 
First, a mesh grid is created to represent the imaged field of view.
Then, the distance between the points of the mesh grid and array elements are calculated based on the known locations of the transducers.
Time of flight is obtained through dividing distance by the SoS values that are assigned based on the temperature of the imaging medium and tissue properties.
The clinical and simulated data are reconstructed with SoS of 1,510\,m/s in this study as the simulations and the experiments were done at the corresponding imaging medium temperature.
Unlike more sophisticated model-based reconstruction approaches~\cite{deanben2012accurate}, backprojection is parameterized using only SoS, making it a stable choice across all imaged scenes.

\subsection{Sparse sampling}
\label{sec:sparse_sampling}

Sparse sampling yields streak artifacts on reconstructed images due to large inter-element pitch size.
For a given angular coverage, i.e.,\ transducer array geometry, using less transducer elements for reconstruction causes stronger artifacts due to increased inter-element pitch size.
We define sparse sampled semi circle array acquisitions \textbf{semi circle sparse 128}, \textbf{semi circle sparse 64} and \textbf{semi circle sparse 32} when using 128, 64 and 32 elements out of the 256 of semi circle array (Fig.~\ref{fig:experimental_data}d first three columns), respectively.
Similarly, we define sparse sampled virtual circle array acquisitions \textbf{virtual circle sparse 128}, \textbf{virtual circle sparse 64} and \textbf{virtual circle sparse 32} when using 128, 64 and 32 elements out of 1,024 of virtual circle array (Fig.~\ref{fig:simulated_data}c first three rows), respectively.
In addition, we also define sparse sampled multisegment array acquisitions \textbf{multisegment sparse 128}, \textbf{multisegment sparse 64} and \textbf{multisegment sparse 32} when using 128, 64 and 32 elements out of the 256 elements of the multisegment array (Figs.~\ref{fig:experimental_data}e and \ref{fig:simulated_data}d first three columns), respectively.
All items correspond to uniform and hence equidistant subsampling of the corresponding transducer array signals.

\subsection{Limited View}
\label{sec:limited_view}

Limited view acquisitions lead to distorted geometry (e.g., elongated vessels) due to the reduced angular coverage (Figs.~\ref{fig:experimental_data}d,e~\&~\ref{fig:simulated_data}c,d last column, limited 128 and linear 128).
To mimic commonly occurring limited view settings, we use a continuous subset of elements for a given transducer.
This corresponds to retaining inter-element pitch size while reducing the angular coverage.
We define a limited view acquisition for each transducer array as follows:
(i) \textbf{Linear array} is the common practice in clinical settings for US data acquisition~\citep{jensen2007medical, deluca2018diagnostic}. 
Typically, the same linear geometry is combined with OA imaging to provide complementary information \citep{mercep2017combined}.
To model this clinically realistic scenario, we use the linear part of the multisegment array for OA image reconstruction.
(ii) \textbf{Semi circle limited 128} uses half of the semi circle array; 128 transducer elements, yielding a quarter circle. 
The differences between linear and semi circle limited view array acquisitions are the inter-element pitch size, focusing and geometry of the active area.
(iii) \textbf{Virtual circle limited 128} uses 128 consecutive elements (one eighth of a circle) out of 1,024.

\section{Datasets}
\label{sec:datasets}

We present four datasets~\footnote{Link to our datasets: \href{https://doi.org/10.3929/ethz-b-000551512}{doi.org/10.3929/ethz-b-000551512}}
~\footnote{Repository for accessing and reading datasets: \href{https://github.com/berkanlafci/oadat}{github.com/berkanlafci/oadat}} 
(two experimental, one simulated, one fully annotated subset) where each has several subcategories for the purpose of tackling different challenges present in the domain. 
Raw signal acquisition data that is used to reconstruct all images are also provided with the datasets.
Experimental datasets also include details about the volunteer Fitzpatrick skin phototype~\citep{gupta2019skin}, which relates to the amount of melanin pigment in the skin (see Appendix \ref{sec:fitzpatrick_skin_phototype_in_experimental_datasets} for distribution and further details).
We also display a comparative overview of publicly available and our proposed OA datasets in Table~\ref{tab:datasets}.
Please refer to Tables~\ref{tab:msfd_contents},~\ref{tab:swfd_sc_contents},~\ref{tab:swfd_ms_contents},~\ref{tab:scd_contents}, and~\ref{tab:oadatmini_contents} for summaries of the file contents of the datasets in the Appendix.

\begin{table}[]
\small
\caption{Publicly available OA datasets, supported tasks, provided data format(s), size, and content.
~\cite{davoudi2019deep} contains 274 mice and 469 phantom slices. \cite{huang2021functional} has 10 mice with 10 frames (100 slices) at 27 different wavelengths and 20 phantom slices.}
\begin{tabular}{@{}lccccccc@{}}
\toprule
\multirow{2}{*}{Dataset} & \multicolumn{3}{c}{tasks}                                                                                                                                                           & \multirow{2}{*}{\begin{tabular}[c]{@{}c@{}}image \\ reconstruction\end{tabular}} & \multirow{2}{*}{\begin{tabular}[c]{@{}c@{}}raw \\ signal\end{tabular}} & \multirow{2}{*}{\begin{tabular}[c]{@{}c@{}}size \\ (\textgreater{}5k instances)\end{tabular}} & \multicolumn{1}{l}{\multirow{2}{*}{\begin{tabular}[c]{@{}l@{}}clinical\\ data\end{tabular}}} \\ \cmidrule(lr){2-4}
                         & \begin{tabular}[c]{@{}c@{}}limited \\ view\end{tabular} & \begin{tabular}[c]{@{}c@{}}sparse \\ sampling\end{tabular} & \begin{tabular}[c]{@{}c@{}}pixel \\ annotations\end{tabular} &                                                                                  &                                                                        &                                                                                            & \multicolumn{1}{l}{}                                                                         \\ \midrule
\cite{davoudi2019deep}                   & \xmark                                                        & \cmark                                                           & \xmark                                                             & \cmark                                                                                  & \xmark                                                                        & \xmark                                                                                     & \xmark                                                                                       \\
\cite{huang2021functional}                   &\xmark                                                         &\xmark                                                            &\xmark                                                              &\xmark                                                                                  &\cmark                                                                        & \xmark                                                                                     & \xmark                                                                                       \\
MSFD (ours)                     & \cmark                                                  & \xmark                                                     & \xmark                                                       & \cmark                                                                           & \cmark                                                                 & \cmark                                                                                     & \cmark                                                                                       \\
SWFD (ours)                     & \cmark                                                  & \cmark                                                     & \xmark                                                       & \cmark                                                                           & \cmark                                                                 & \cmark                                                                                     & \cmark                                                                                       \\
SCD (ours)                      & \cmark                                                  & \cmark                                                     & \cmark                                                       & \cmark                                                                           & \cmark                                                                 & \cmark                                                                                     & \xmark    
\\
OADAT-mini (ours)                      & \cmark                                                  & \cmark                                                     & \cmark                                                       & \cmark                                                                           & \cmark                                                                 & \cmark                                                                                     & \xmark ~$\cup$ \cmark   
\\ \bottomrule
\end{tabular}
\label{tab:datasets}
\end{table}

\subsection{Multispectral forearm dataset}
\label{sec:msfd}

Multispectral forearm dataset (MSFD) is collected using multisegment array (Sec.~\ref{sec:transducer_arrays}) from nine volunteers at six different wavelengths (700, 730, 760, 780, 800, 850\,nm) for both arms. 
Selected wavelengths are particularly aimed for spectral decomposition aiming to separate oxy- and deoxy-hemoglobin~\citep{perekatova2016optimal}.
All wavelengths are acquired consecutively, yielding almost identical scene being captured for a given slice across different wavelengths with slight displacement errors. 
For each of the mentioned category 1,400 slices are captured, creating a sum of $9 \times 6 \times 2 \times 1,400 = 151,200$ unique signal matrices. 

From this data, using backprojection algorithm, we reconstruct (i) linear array images $\mathrm{MSFD_{linear}}$, (ii) multisegment array images $\mathrm{MSFD_{ms}}$,  
(iii) multisegment sparse 128 images (Sec.\ref{sec:sparse_sampling}), $\mathrm{MSFD_{ms, ss128}}$,
(iv) multisegment sparse 64 images (Sec.\ref{sec:sparse_sampling}), $\mathrm{MSFD_{ms, ss64}}$, and
(v) multisegment sparse 32 images (Sec.\ref{sec:sparse_sampling}), $\mathrm{MSFD_{ms, ss32}}$,
where each dataset has 151,200 images of $256 \times 256$ pixel resolution; totaling to 756,000 image instances.

\subsection{Single wavelength forearm dataset}
\label{sec:swfd}

Single wavelength forearm dataset (SWFD) is collected using both multisegment and semi circle arrays (Sec.~\ref{sec:transducer_arrays}) from 14 volunteers at a single wavelength (1,064\,nm) for both arms. 
The choice of the wavelength is based on maximizing penetration depth for excitation light source (laser)~\citep{sharma2019photoacoustic}.
Out of the 14 volunteers, eight of them have also participated in the MSFD experiment and their unique identifiers match across the dataset files. 
For each array, volunteer, and arm, we acquired 1,400 slices, creating a sum of $2 \times 14 \times 2 \times 1,400 = 78,400$ unique signal matrices. 
It is important to note that despite the data being acquired from the same volunteers, signals between multisegment array and semi circle array are not paired due to physical constraints.

From this data, using backprojection algorithm, we reconstruct (i) linear array images, $\mathrm{SWFD_{linear}}$,  (ii) multisegment array images, $\mathrm{SWFD_{ms}}$, (iii) semi circle array images, $\mathrm{SWFD_{sc}}$, (iv) semi circle array limited 128 images (Sec.\ref{sec:limited_view}), $\mathrm{SWFD_{sc, lv128}}$, (v) semi circle sparse 128 images (Sec.\ref{sec:sparse_sampling}), $\mathrm{SWFD_{sc, ss128}}$, (vi) semi circle sparse 64 images (Sec.\ref{sec:sparse_sampling}), $\mathrm{SWFD_{sc, ss64}}$, (vii) semi circle sparse 32 images (Sec.\ref{sec:sparse_sampling}), $\mathrm{SWFD_{sc, ss32}}$, 
(viii) multisegment sparse 128 images (Sec.\ref{sec:sparse_sampling}), $\mathrm{SWFD_{ms, ss128}}$, 
(ix) multisegment sparse 64 images (Sec.\ref{sec:sparse_sampling}), $\mathrm{SWFD_{ms, ss64}}$, and 
(x) multisegment sparse 32 images (Sec.\ref{sec:sparse_sampling}), $\mathrm{SWFD_{ms, ss32}}$,
where each dataset has 39,200 images of $256 \times 256$ pixel resolution; totaling to 392,000 image instances. 

\subsection{Simulated cylinders dataset}
\label{sec:scd}

Simulated cylinders dataset (SCD) is a group of synthetically generated 20,000 forearm acoustic pressure maps that we heuristically produced based on a range of criteria we observed in experimental images.
The acoustic pressure maps are generated with $256 \times 256$ pixel resolution where skin curves and afterwards a random amount of ellipses with different intensity profiles are generated iteratively for a given image (see Fig.~\ref{fig:simulated_data}). 
We explain details for the simulation algorithm\footnote{Python module for acoustic map simulation: \href{https://renkulab.io/gitlab/firat.ozdemir/oa-armsim}{renkulab.io/gitlab/firat.ozdemir/oa-armsim}.} 
for generating acoustic pressure map in Appendix \ref{sec:simulated_cylinder_dataset_generation}. 

Based on the acoustic pressure map, we generate its annotation map with three labels, corresponding to background, vessels, and skin curve. 
For each acoustic pressure map, we generate signal matrices for the geometries of linear, multisegment and virtual circle arrays. 
Using linear and multisegment array signals, we use backprojection algorithm to reconstruct (i) linear array images, $\mathrm{SCD_{linear}}$, and (ii) multisegment array images, $\mathrm{SCD_{ms}}$, (iii) multisegment sparse 128 images (Sec.\ref{sec:sparse_sampling}), $\mathrm{SCD_{ms, ss128}}$, (iv) multisegment sparse 64 images (Sec.\ref{sec:sparse_sampling}), $\mathrm{SCD_{ms, ss64}}$, and (v) multisegment sparse 32 images (Sec.\ref{sec:sparse_sampling}), $\mathrm{SCD_{ms, ss32}}$.
From virtual circle array signals, we use backprojection algorithm to reconstruct (vi) virtual circle images, $\mathrm{SCD_{vc}}$, (vii) virtual circle limited 128 images  (Sec.\ref{sec:limited_view}), $\mathrm{SCD_{vc, lv128}}$, (viii) virtual circle sparse 128 images (Sec.\ref{sec:sparse_sampling}), $\mathrm{SCD_{vc, ss128}}$, (ix) virtual circle sparse 64 images (Sec.\ref{sec:sparse_sampling}), $\mathrm{SCD_{vc, ss64}}$, and (x) virtual circle sparse 32 images (Sec.\ref{sec:sparse_sampling}), $\mathrm{SCD_{vc, ss32}}$, where each dataset has 20,000 images of $256 \times 256$ pixel resolution; totaling to 200,000 image instances. 
All ten image reconstruction variations of SCD have corresponding pairs for each of the 20k image; i.e., produced from the same acoustic pressure map. 

\subsection{OADAT-mini dataset}
Using a subset of 100 signal and corresponding reconstruction instances from each of the previously mentioned datasets, we also present OADAT-mini, which is a fragment of OADAT that is significantly smaller yet comprehensive for understanding the contents of OADAT.
In addition, OADAT-mini contains manual annotation maps for vessels in the reconstructed images.

\section{Tasks}
\label{sec:tasks}

\begin{table}[]
\centering
\footnotesize
\caption{List of tasks and experiments we define on MSFD, SWFD, and SCD.
Experiment names are built of (i) dataset name for translation tasks or seg for segmentation task, (ii) input data and corresponding number of active array elements; sparse sampling (ss), limited view (lv), virtual circle (vc), and (iii) input array type; semi circle (sc), virtual circle (vc), linear (li), multisegment (ms). 
Image data and annotation maps are represented with $x$ and $y$, while predicted image and annotations are shown as $x^*$ and $y^*$.}
\begin{tabular}{@{}ll@{}}
\toprule
\textbf{Limited view correction}                                                                                                                                          & \textbf{Semantic segmentation}                                                                       \\ \midrule
$f_{\mathrm{MSFD\_lv128, li}}$: $x \sim \mathrm{MSFD_{linear}} \to x^* \sim \mathrm{MSFD_{ms}}$                                                                                            & $f_{\mathrm{seg\_vc, vc}}$: $x \sim \mathrm{SCD_{vc}} \to y^* \sim \mathrm{labels}$                                   \\
$f_{\mathrm{SWFD\_lv128, li}}$: $x \sim \mathrm{SWFD_{linear}} \to x^* \sim \mathrm{SWFD_{ms}}$                                                                                            & $f_{\mathrm{seg\_lv128, li}}$: $x \sim \mathrm{SCD_{linear}} \to y^* \sim \mathrm{labels}$                            \\
$f_{\mathrm{SWFD\_lv128, sc}}$: $x \sim \mathrm{SWFD_{sc, lv128}} \to x^* \sim \mathrm{SWFD_{sc}}$                                                                                         & $f_{\mathrm{seg\_lv128, vc}}$: $x \sim \mathrm{SCD_{vc, lv128}} \to y^* \sim \mathrm{labels}$                         \\
$f_{\mathrm{SCD\_lv128, li}}$: $x \sim \mathrm{SCD_{linear}} \to x^* \sim \mathrm{SCD_{vc}}$                                                                                               & $f_{\mathrm{seg\_ss128, vc}}$: $x \sim \mathrm{SCD_{vc, ss128}} \to y^* \sim \mathrm{labels}$                         \\
$f_{\mathrm{SCD\_lv128, vc}}$: $x \sim \mathrm{SCD_{vc, lv128}} \to x^* \sim \mathrm{SCD_{vc}}$                                                                                            & $f_{\mathrm{seg\_ss64, vc}}$: $x \sim \mathrm{SCD_{vc, ss64}} \to y^* \sim \mathrm{labels}$                           \\
$f_{\mathrm{SCD\_lv256, ms}}$: $x \sim \mathrm{SCD_{ms}} \to x^* \sim \mathrm{SCD_{vc}}$                                                                                                   & $f_{\mathrm{seg\_ss32, vc}}$: $x \sim \mathrm{SCD_{vc, ss32}} \to y^* \sim \mathrm{labels}$                           \\ \cmidrule(r){1-1}
\textbf{Sparse sampling correction}                                                                                                                                       & \smash{\raisebox{0.65em}{$f_{\mathrm{seg\_ss128, ms}}$: $x \sim \mathrm{SCD_{ms, ss128}} \to y^* \sim \mathrm{labels}$}} \\
$f_{\mathrm{SWFD\_ss128, sc}}$: $x \sim \mathrm{SWFD_{sc, ss128}} \to x^* \sim \mathrm{SWFD_{sc}}$                                                                                         & \smash{\raisebox{0.65em}{$f_{\mathrm{seg\_ss64, ms}}$: $x \sim \mathrm{SCD_{ms, ss64}} \to y^* \sim \mathrm{labels}$}}   \\
$f_{\mathrm{SWFD\_ss64, sc}}$: $x \sim \mathrm{SWFD_{sc, ss64}} \to x^* \sim \mathrm{SWFD_{sc}}$                                                                                           & \smash{\raisebox{0.65em}{$f_{\mathrm{seg\_ss32, ms}}$: $x \sim \mathrm{SCD_{ms, ss32}} \to y^* \sim \mathrm{labels}$} }  \\
$f_{\mathrm{SWFD\_ss32, sc}}$: $x \sim \mathrm{SWFD_{sc, ss32}} \to x^* \sim \mathrm{SWFD_{sc}}$		& \smash{\raisebox{0.35em}{$f_{\mathrm{seg\_MSFD\_lv128, ms}}$: $x \sim \mathrm{MSFD_{linear}} \to y^* \sim \mathrm{labels}$} }  \\
$f_{\mathrm{SCD\_ss128, vc}}$: $x \sim \mathrm{SCD_{vc, ss128}} \to x^* \sim \mathrm{SCD_{vc}}$			& \smash{\raisebox{0.35em}{$f_{\mathrm{seg\_MSFD\_ss128, ms}}$: $x \sim \mathrm{MSFD_{ms, ss128}} \to y^* \sim \mathrm{labels}$} }  \\
$f_{\mathrm{SCD\_ss64, vc}}$: $x \sim \mathrm{SCD_{vc, ss64}} \to x^* \sim \mathrm{SCD_{vc}}$			& \smash{\raisebox{0.35em}{$f_{\mathrm{seg\_MSFD\_ss64, ms}}$: $x \sim \mathrm{MSFD_{ms, ss64}} \to y^* \sim \mathrm{labels}$} } \\
$f_{\mathrm{SCD\_ss32, vc}}$: $x \sim \mathrm{SCD_{vc, ss32}} \to x^* \sim \mathrm{SCD_{vc}}$			& \smash{\raisebox{0.35em}{$f_{\mathrm{seg\_MSFD\_ss32, ms}}$: $x \sim \mathrm{MSFD_{ms, ss32}} \to y^* \sim \mathrm{labels}$} }\\
$f_{\mathrm{SWFD\_ss128, ms}}$: $x \sim \mathrm{SWFD_{ms, ss128}} \to x^* \sim \mathrm{SWFD_{ms}}$ &    \smash{\raisebox{0.35em}{$f_{\mathrm{seg\_SWFD\_lv128, ms}}$: $x \sim \mathrm{SWFD_{linear}} \to y^* \sim \mathrm{labels}$} } \\
$f_{\mathrm{SWFD\_ss64, ms}}$: $x \sim \mathrm{SWFD_{ms, ss64}} \to x^* \sim \mathrm{SWFD_{ms}}$    &   \smash{\raisebox{0.35em}{$f_{\mathrm{seg\_SWFD\_lv128, sc}}$: $x \sim \mathrm{SWFD_{sc, lv128}} \to y^* \sim \mathrm{labels}$} } \\
$f_{\mathrm{SWFD\_ss32, ms}}$: $x \sim \mathrm{SWFD_{ms, ss32}} \to x^* \sim \mathrm{SWFD_{ms}}$    &   \smash{\raisebox{0.35em}{$f_{\mathrm{seg\_SWFD\_ms, ms}}$: $x \sim \mathrm{SWFD_{ms}} \to y^* \sim \mathrm{labels}$} }\\
$f_{\mathrm{SCD\_ss128, ms}}$: $x \sim \mathrm{SCD_{ms, ss128}} \to x^* \sim \mathrm{SCD_{vc}}$     &	\smash{\raisebox{0.35em}{$f_{\mathrm{seg\_SWFD\_sc}}$: $x \sim \mathrm{SWFD_{sc}} \to y^* \sim \mathrm{labels}$} }\\
$f_{\mathrm{SCD\_ss64, ms}}$: $x \sim \mathrm{SCD_{ms, ss64}} \to x^* \sim \mathrm{SCD_{vc}}$       &    \smash{\raisebox{0.35em}{$f_{\mathrm{seg\_SWFD\_ss128, sc}}$: $x \sim \mathrm{SWFD_{sc, ss128}} \to y^* \sim \mathrm{labels}$} }\\
$f_{\mathrm{SCD\_ss32, ms}}$: $x \sim \mathrm{SCD_{ms, ss32}} \to x^* \sim \mathrm{SCD_{vc}}$       &    \smash{\raisebox{0.35em}{$f_{\mathrm{seg\_SWFD\_ss64, sc}}$: $x \sim \mathrm{SWFD_{sc, ss64}} \to y^* \sim \mathrm{labels}$} }\\
$f_{\mathrm{MSFD\_ss128, ms}}$: $x \sim \mathrm{MSFD_{ms, ss128}} \to x^* \sim \mathrm{MSFD_{ms}}$  &    \smash{\raisebox{0.35em}{$f_{\mathrm{seg\_SWFD\_ss32, sc}}$: $x \sim \mathrm{SWFD_{sc, ss32}} \to y^* \sim \mathrm{labels}$} }\\
$f_{\mathrm{MSFD\_ss64, ms}}$: $x \sim \mathrm{MSFD_{ms, ss64}} \to x^* \sim \mathrm{MSFD_{ms}}$    &    \smash{\raisebox{0.35em}{$f_{\mathrm{seg\_SWFD\_ss128, ms}}$: $x \sim \mathrm{SWFD_{ms, ss128}} \to y^* \sim \mathrm{labels}$} }\\
$f_{\mathrm{MSFD\_ss32, ms}}$: $x \sim \mathrm{MSFD_{ms, ss32}} \to x^* \sim \mathrm{MSFD_{ms}}$    &    \smash{\raisebox{0.35em}{$f_{\mathrm{seg\_SWFD\_ss64, ms}}$: $x \sim \mathrm{SWFD_{ms, ss64}} \to y^* \sim \mathrm{labels}$} }\\
&	\smash{\raisebox{0.35em}{$f_{\mathrm{seg\_SWFD\_ss32, ms}}$: $x \sim \mathrm{SWFD_{ms, ss32}} \to y^* \sim \mathrm{labels}$} } \\
\bottomrule
\end{tabular}
\label{tab:tasks}
\end{table}

Based on the datasets presented in Sec.~\ref{sec:datasets}, we define a list of experiments related to image translation to overcome (i) sparse sampling and (ii) limited view artifacts, and semantic segmentation of images. 

\subsection{Image translation}
\label{sec:image_translation}

Through a list of permutations of our datasets, we can define several pairs of varying difficulty of image translation experiments where the target images are also available (see Table~\ref{tab:tasks}).
We present sparse sampling and limited view reconstructions of SWFD, MSFD and SCD for all transducer arrays.
Sparse sampling correction experiments learn mapping functions listed in Table~\ref{tab:tasks}, where the function notations denote the dataset used, the task of sparse sampling (ss) correction from the given number of elements used for image reconstruction and the array that is used to generate the input.  
Limited view correction experiments learn mapping functions, listed in Table~\ref{tab:tasks}, where the function notations denote the dataset used, the task of limited view (lv) correction from the given number of elements used for image reconstruction and the array that is used to generate the input. 

\subsection{Semantic segmentation}

SCD includes pixel annotations for skin curve, vessels and background.
In addition to segmentation of these structures on the ideal reconstructions $\mathrm{SCD_{vc}}$, we define this task on sparse sampling and limited view reconstructions that contain the relevant artifacts encountered in experimental data. 
Accordingly, we compose the nine segmentation experiments listed in Table~\ref{tab:tasks}, where the function notations denote the task segmentation (seg), type of the reconstructed input being used (virtual circle (vc), limited view (lv), and sparse sampling (ss)) and the array that is used to generate the input.
All data is generated from SCD and the objective is to match the ground truth annotations of the acoustic pressure map. 
Different than SCD, experimental datasets under OADAT-mini include pixel annotations for vessels and consist of 14 segmentation experiments, also listed in Table~\ref{tab:tasks}. 

\section{Experiments and results}

\begin{wrapfigure}{5}{0.4\textwidth}
\vspace{-4.em}
  \begin{center}
  \centering
  \includegraphics[width=0.4\textwidth]{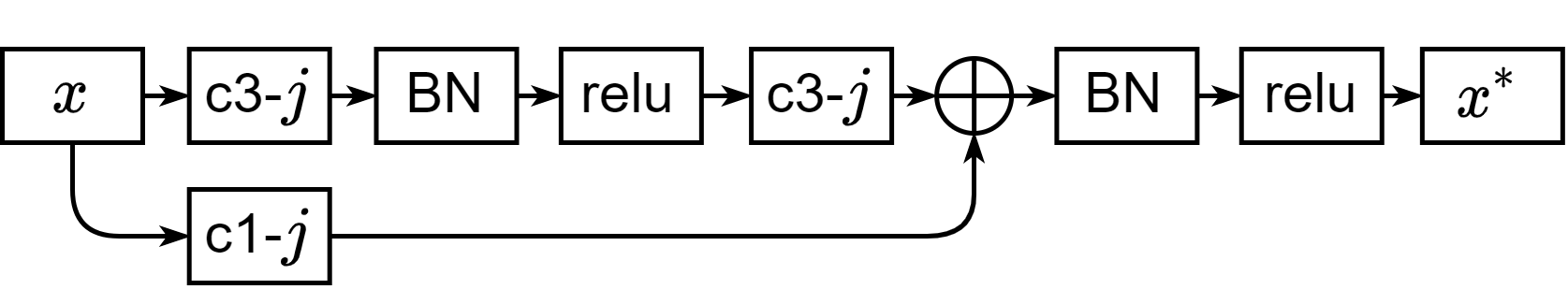}
  \end{center}
  \caption{Residual convolutional block with batch normalization (BN). 
  c$i$-$j$ conv.\,layer have $i \times i$ kernels and $j$ filters.}
  \label{fig:resconvblockBN}
\end{wrapfigure}

For all experiments we standardize the architecture that we based on UNet~\citep{ronneberger2015u}. 
Specifically, we adopt the five spatial feature abstraction levels and use skip connections to concatenate with features of matching spatial dimension along the upsampling path. 
However, we make several distinct design choices that vary from vanilla UNet. 
First, we use attention gates~\citep{oktay2018attention} at the end of each skip connection.
Second, we opt for residual convolutional blocks with batch normalization~\citep{ioffe2015batch} at each level, shown in Fig.~\ref{fig:resconvblockBN}.
Third, we use 2D bilinear upsampling instead of deconvolutions. 
Finally, we use half the number of convolutional kernels at each layer; e.g., start with 32 convolutional filters as opposed to 64. 
Full schematic as well as other implementation details are discussed in Appendix \ref{sec:architecture_and_implementation_details}. 
We refer to our modified UNet architecture as modUNet hereon.

\subsection{Data split and preprocessing}

We standardize how we split each dataset into training and test sets regardless of the task in order to ensure consistency in our and future experiments. 
Out of the nine volunteers in MSFD, we use five for training (IDs: 2, 5, 6, 7, 9), one for validation (ID: 10) and three for testing (IDs: 11, 14, 15). 
Out of the 14 volunteers in SWFD, we use eight for training (IDs: 1, 2, 3, 4, 5, 6, 7, 8), one for validation (ID: 9) and five for testing (IDs: 10, 11, 12, 13, 14). 
Out of the 20k slices in SCD, we use the first 14k for training, following 1k for validation, and the last 5k for testing. 
For each experiment conducted on OADAT-mini, we use the first 75 samples for training, next 5 for validation and the last 20 for quantitative evaluation.
This translates to six times the numbers for MSFD-mini, where there are 100 samples for each of the six wavelengths. 
As before, we conduct the MSFD-mini segmentation experiments as a single modUNet attempting segment any of the given six wavelength samples. 
As a preprocessing step, all data instances (except for annotation maps) are independently scaled by their maximum and then clipped at a minimum value of $-0.2$ \citep{ding2015efficient}.

\subsection{Results}
\label{sec:results}

\begin{table}[!ht]
\small
\centering
\caption{Image translation results of the proposed modUNet model reported as mean \textpm std.
Each row corresponds to the results of the experiment where input data is identified through (i) the input data and corresponding number of active transducer elements; sparse sampling (ss), limited view (lv) and (ii) the array type used for input; semi circle (sc), virtual circle (vc), linear (li), multisegment (ms). }

\begin{tabular}{@{}llllll@{}}
\toprule
                              &          & \multicolumn{1}{c}{MAE} & \multicolumn{1}{c}{RMSE} & \multicolumn{1}{c}{SSIM} & \multicolumn{1}{c}{PSNR} \\ \midrule
\textbf{SCD}                  &          &                         &                          &                          &                          \\
\multirow{3}{*}{Limited view} & lv128,li & 0.005 \textpm 1.3e-3          & 0.012 \textpm 3.3e-3           & 0.978 \textpm 7.2e-3           & 39.909 \textpm 2.15            \\
                              & lv128,vc & 0.005 \textpm 1.3e-3          & 0.013 \textpm 3.4e-3           & 0.978 \textpm 6.2e-3           & 39.853 \textpm 2.18            \\
                              & lv256,ms & 0.005 \textpm 1.2e-3          & 0.009 \textpm 3.0e-3           & 0.984 \textpm 6.1e-3           & 42.459 \textpm 2.52            \\[1.5ex] 
\multirow{6}{*}{Sparse view}  & ss128,vc & 0.004 \textpm 1.1e-3          & 0.008 \textpm 2.0e-3           & 0.985 \textpm 6.1e-3           & 44.068 \textpm 2.12            \\
                              & ss64,vc  & 0.005 \textpm 1.3e-3          & 0.011 \textpm 2.9e-3           & 0.981 \textpm 6.9e-3          & 41.402 \textpm 2.25            \\
                              & ss32,vc  & 0.006 \textpm 1.5e-3          & 0.013 \textpm 3.7e-3           & 0.973 \textpm 8.7e-3           & 39.309 \textpm 2.23            \\
                               & ss128,ms & 0.005 \textpm 1.1e-3          & 0.010 \textpm 3.0e-3           & 0.984 \textpm 6.1e-3           & 42.345 \textpm 2.39            \\
                              & ss64,ms  & 0.005 \textpm 1.2e-3          & 0.010 \textpm 3.0e-3           & 0.982 \textpm 6.7e-3           & 41.525 \textpm 2.19            \\
                              & ss32,ms & 0.005 \textpm 1.3e-3          & 0.011 \textpm 3.0e-3           & 0.979 \textpm 7.3e-3           & 40.678 \textpm 2.10            \\ \midrule
\textbf{SWFD}                 &          &                         &                          &                          &                          \\
\multirow{2}{*}{Limited view} & lv128,li & 0.028 \textpm 1.6e-2          & 0.039 \textpm 2.0e-2           & 0.613 \textpm 1.4e-1           & 29.397 \textpm 4.75            \\
                              & lv128,sc & 0.016 \textpm 1.2e-2          & 0.021 \textpm 1.5e-2           & 0.811 \textpm 1.3e-1          & 36.791 \textpm 5.30            \\[1.5ex]
\multirow{6}{*}{Sparse view}  & ss128,sc & 0.015 \textpm 1.2e-2          & 0.019 \textpm 1.5e-2           & 0.863 \textpm 1.0e-1           & 38.233 \textpm 5.35            \\
                              & ss64,sc  & 0.019 \textpm 1.5e-2          & 0.024 \textpm 1.8e-2           & 0.769 \textpm 1.6e-1          & 35.605 \textpm 5.22            \\
                              & ss32,sc  & 0.021 \textpm 1.6e-2          & 0.028 \textpm 2.0e-2           & 0.693 \textpm 2.0e-1           & 33.852 \textpm 5.37           \\
                              & ss128,ms & 0.023 \textpm 1.5e-2          & 0.028 \textpm 1.9e-2           & 0.784 \textpm 9.7e-2          & 33.764 \textpm 4.53           \\
                              & ss64,ms  & 0.029 \textpm 1.9e-2          & 0.037 \textpm 2.3e-2           & 0.636 \textpm 1.5e-1           & 31.311 \textpm 4.80            \\
                              & ss32,ms  & 0.033 \textpm 2.1e-2         & 0.042 \textpm 2.5e-2           & 0.521 \textpm 1.8e-1           & 29.813 \textpm 5.11           \\ \midrule
\textbf{MSFD}                 &          &                         &                          &                          &                          \\
limited view                  & lv128,li & 0.023 \textpm 1.1e-2          & 0.035 \textpm 1.4e-2           & 0.640 \textpm 1.4e-1           & 29.731 \textpm 3.97           \\ [1.5ex]
\multirow{3}{*}{Sparse view}  & ss128,ms & 0.017 \textpm 9.7e-3          & 0.022 \textpm 1.2e-2           & 0.839 \textpm 8.1e-2           & 35.798 \textpm 3.84            \\
                              & ss64,ms  & 0.022 \textpm 1.2e-2          & 0.029 \textpm 1.5e-2        & 0.719 \textpm 1.4e-1           & 33.104 \textpm 4.02            \\
                              & ss32,ms & 0.026 \textpm 1.4e-2          & 0.036 \textpm 1.7e-2           & 0.608 \textpm 1.8e-1           & 30.873 \textpm 4.22            \\\bottomrule
\end{tabular}
\label{tab:translation}
\end{table}

\begin{figure}[!b]
\centering
    \begin{subfigure}[b]{0.43\linewidth}%
    \centering%
        \includegraphics[width=\textwidth, trim= 0cm 0.0cm 0cm 0cm, clip]{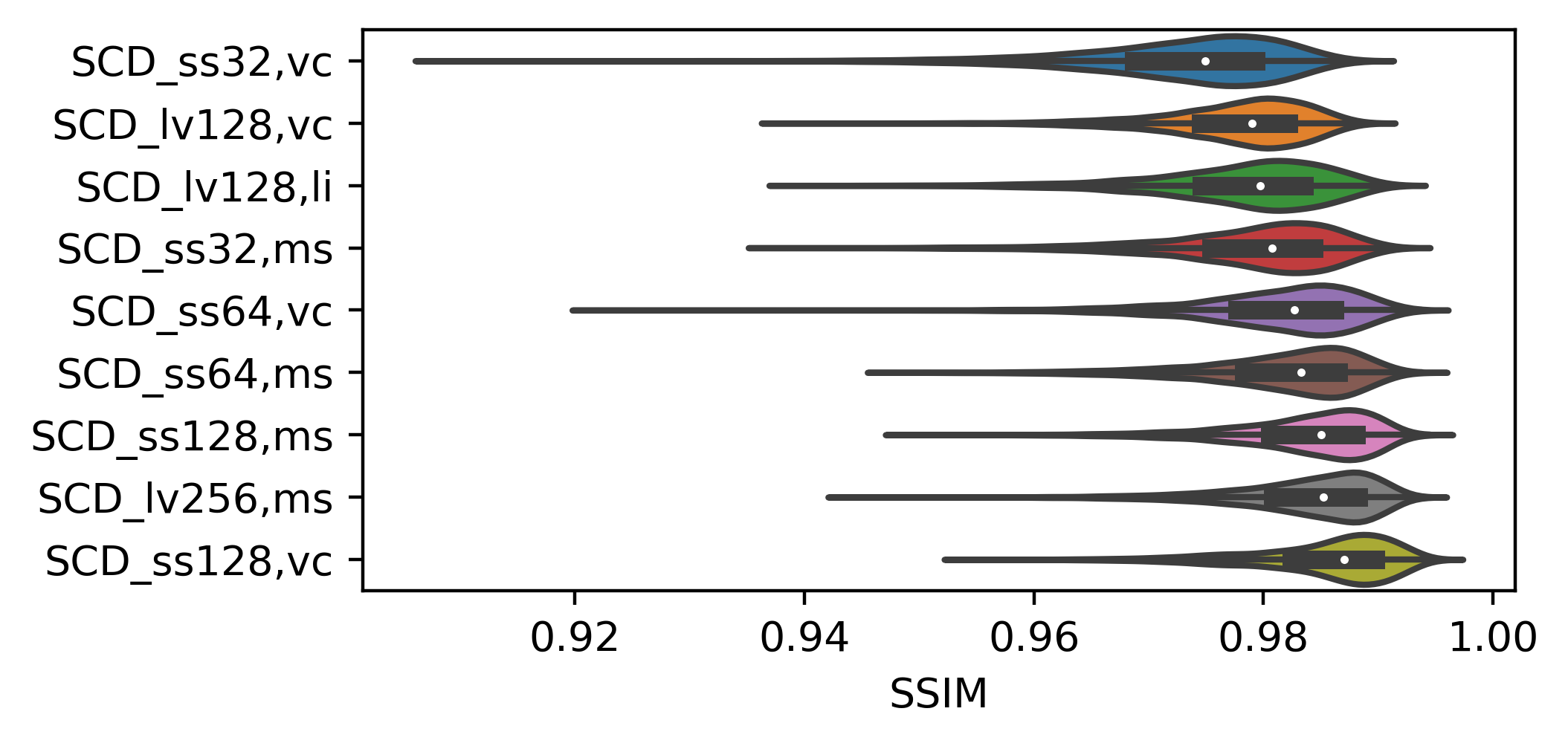}
    \end{subfigure}%
	\begin{subfigure}[b]{0.3\linewidth}%
    \centering
        \includegraphics[width=\textwidth, trim= 0cm 0.0cm 0cm 0cm, clip]{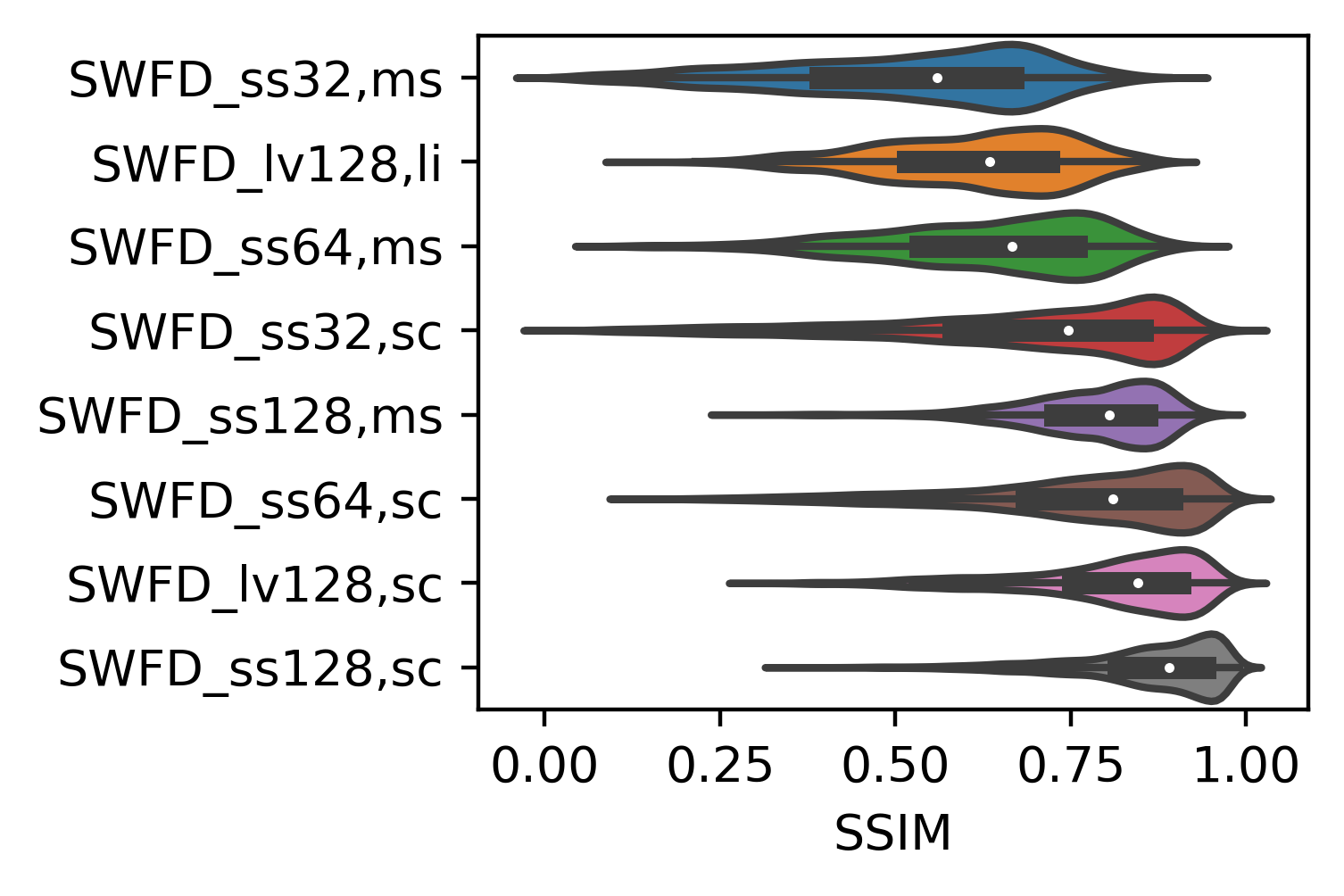}
    \end{subfigure}%
    \begin{subfigure}[b]{0.29\linewidth}%
    \centering%
        \includegraphics[width=\textwidth, trim= 0cm 0.0cm 0cm 0cm, clip]{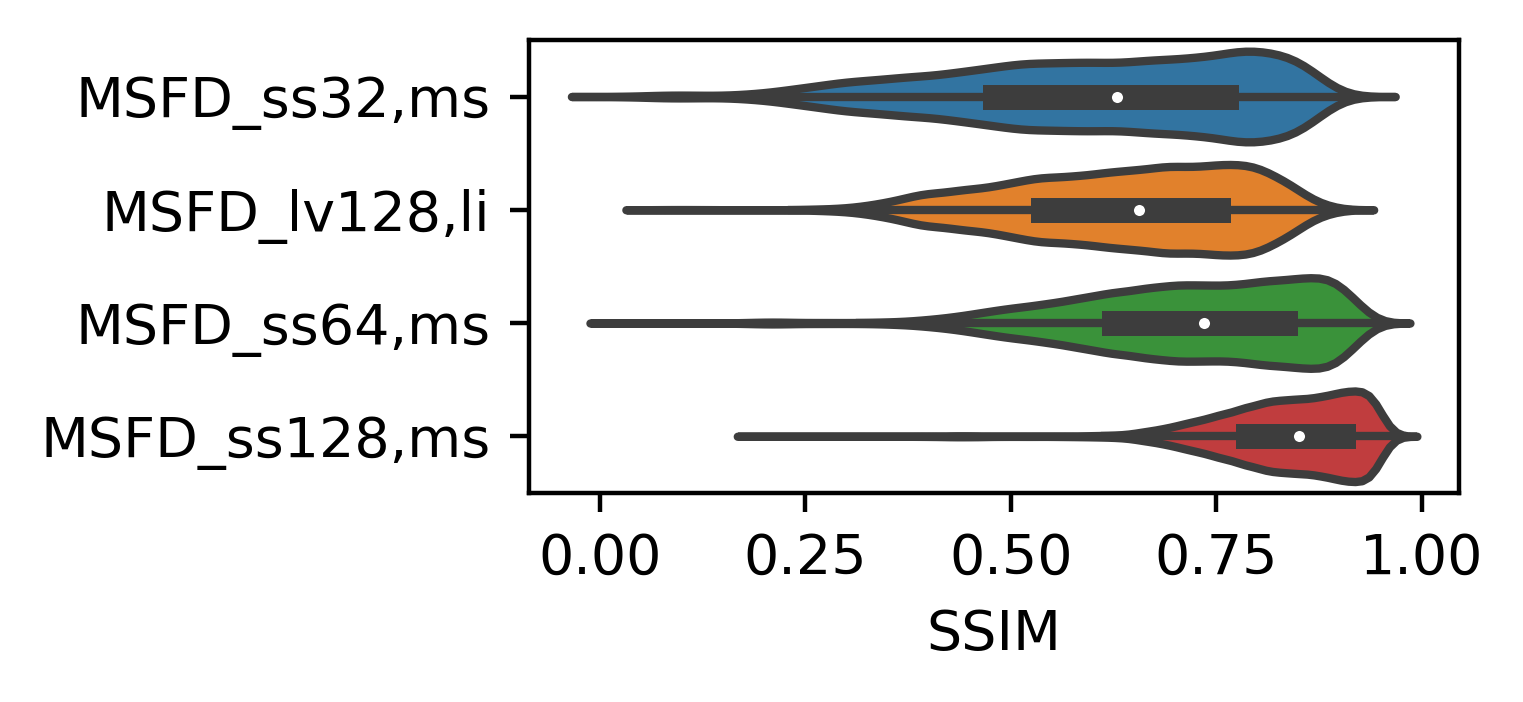}
    \end{subfigure}%
    \caption{Distribution of modUNet SSIM performance on SCD(left), SWFD (middle) and MSFD (right) image translation experiments, sorted in ascending median test sample performance.}
    \label{fig:image_translation}
\end{figure}

We evaluate modUNet performance on the test sets using standard metrics.
Namely, we report mean absolute error (MAE), root mean squared error (RMSE), structural similarity index (SSIM), and peak signal-to-noise ratio (PSNR) for image translation experiments between modUNet predictions and targets. 
Segmentation task performance is reported using Dice coefficient (F1-score), intersection over union (IoU, i.e., Jaccard index) and 95-percentile Hausdorff distance (HD95) metrics between modUNet predictions and annotation maps for vessels and skin curves.
HD95 is calculated in several steps: 
First, the set of pixels along the contour for each predicted (set A) and annotated (set B) target structures are found.
For each set point, the closest point from the other set is determined based on $l_2$ distance.
Different from the standard Hausdorff distance, the 95-percentile distance value is taken as the directional distance from set A to B (and vice versa) instead of the maximum distance.
Then the maximum of these two values is calculated as the symmetric HD95 for a given image.
In OADAT-mini experiments, only vessel annotation maps are available.
Since OADAT-mini consist of subsets of the other three datasets, we do not repeat image translation tasks.
In Tables~\ref{tab:translation}~\&~\ref{tab:segmentation} we report modUNet results for mean and standard deviations aggregated over the corresponding test set images. 

Using SSIM, we show performance across all our datasets in Fig.~\ref{fig:image_translation}.
Upon exploring the reason behind the long tails, we notice that most of the lower scores occur when acquisition noise and/or artifacts are more pronounced.
Depending on the sample, this can imply either modUNet reduced the noise present in the target, or both input/output pair in the test set had low signal-to-noise ratio(SNR).
Nevertheless, modUNet successfully corrects geometric distortions for limited view experiments.
Given the low mean and standard deviations in MAE and RMSE, we can comment that modUNet can generalize well to previously unseen volunteer data.
This is further corroborated with the narrow SSIM interquartile range in Fig.~\ref{fig:image_translation} violin plots.

\begin{figure}
\centering
	\begin{subfigure}[b]{0.496\linewidth}
    \centering
        \includegraphics[width=\textwidth, trim= 0cm 0.0cm 0cm 0cm, clip]{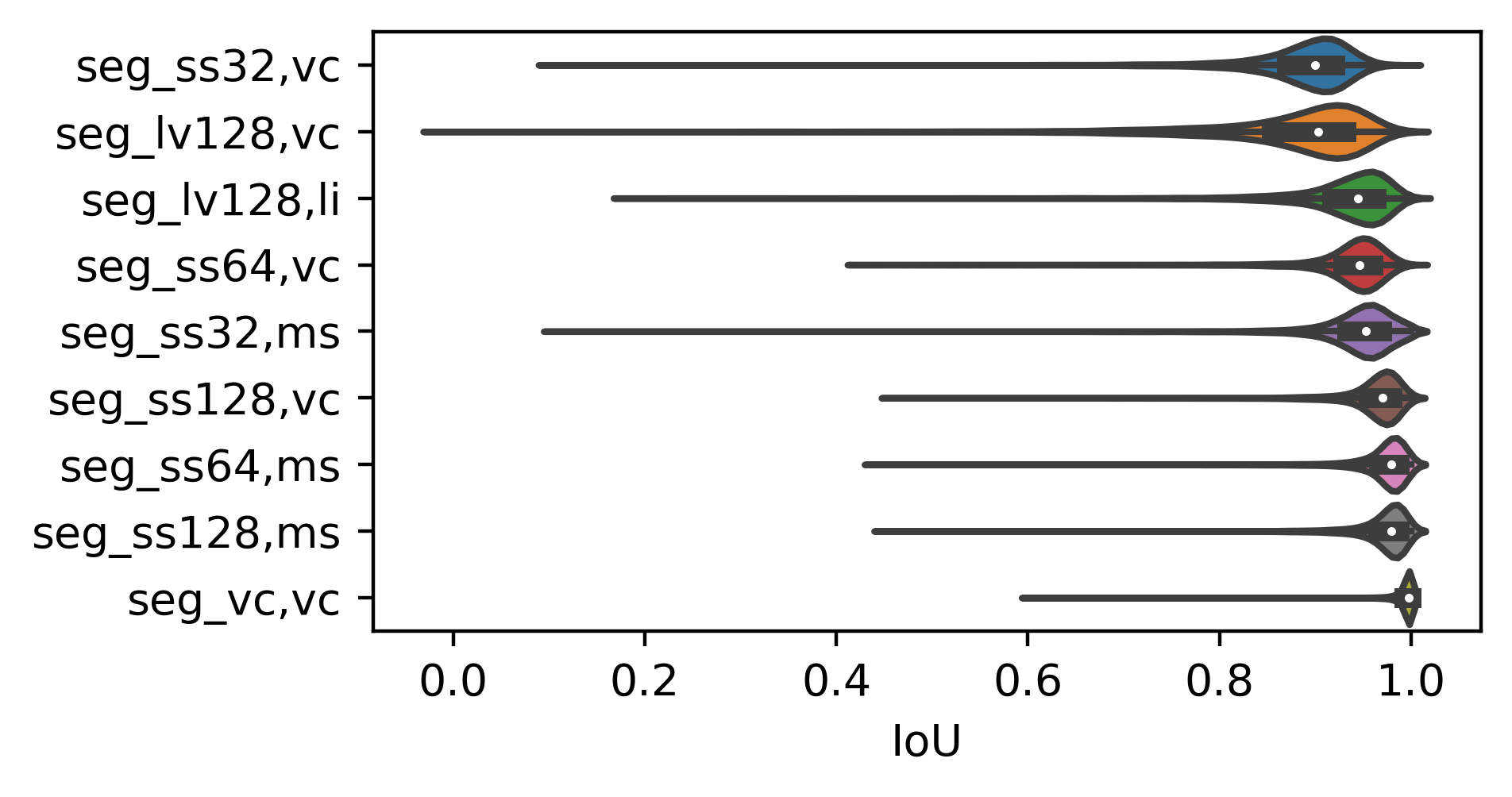}
    \end{subfigure}
    \begin{subfigure}[b]{0.496\linewidth}
    \centering
        \includegraphics[width=\textwidth, trim= 0cm 0.0cm 0cm 0cm, clip]{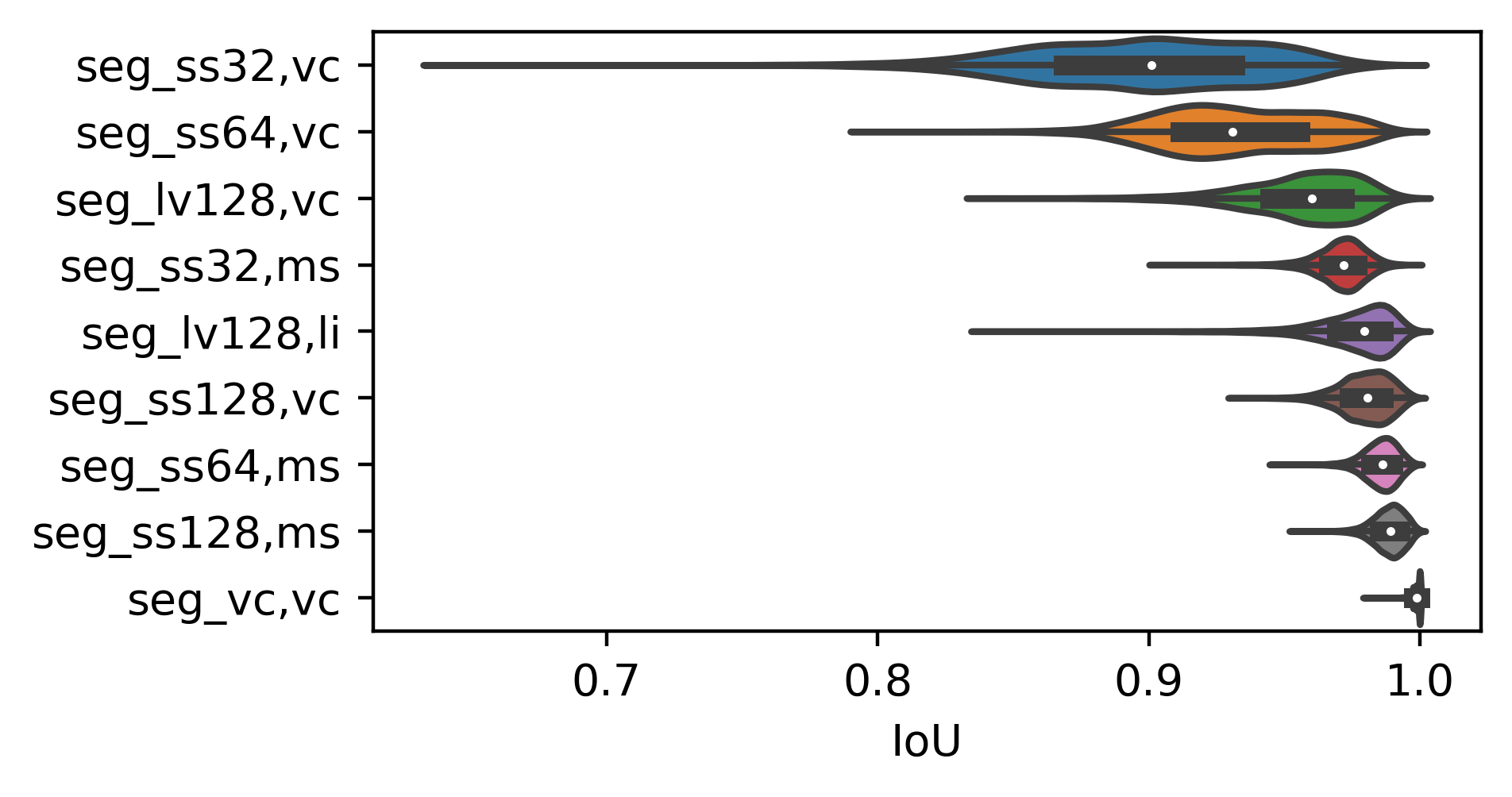}
    \end{subfigure}\hfill
    \caption{Distribution of modUNet IoU performance on SCD semantic segmentation experiments for vessel (left) and skin curve (right) labels, sorted in ascending median test sample performance.}
    \label{fig:semantic_segmentation}
\end{figure}

Similarly, we plot the segmentation performance for IoU across different experiments on SCD in Fig.~\ref{fig:semantic_segmentation} for vessel and skin curve labels.  
While skin curve segmentation performance almost never drops below IoU of 0.80 for SCD, one can see that IoU can be drastically lower for vessel segmentation. 
We observed that this only happens when the size of vessel is small.
For example, there are examples with ground truth vessels having as low as four pixels while the prediction has six, leading to an IoU of $0.6$.
As for experiments with OADAT-mini, vessel segmentation performance in experimental datasets are worse.
Specifically, the worst scores are observed for MSFD-mini experiments.
This is due to the movement artifact between different wavelengths of a given slice.
All MSFD-mini instances have expert annotations for 800\,nm wavelength. 
Slight movement across different wavelengths can yield poor quantitative metrics, particularly exacerbated when the observed vessels are small.
Given the limited training and test sizes of OADAT-mini experiments, we believe that the quantitative results should be taken as a reference.
The qualitative results in the Appendix can be more informative for gaining insight for modUNet performance when trained with a very limited amount of data.
We provide qualitative results as well as conduct further analysis for all tasks in Appendix~\ref{sec:qualitative_results}.

\begin{table}[]
\scriptsize
\centering
\caption{Segmentation results of our proposed modUNet model reported as mean \textpm std.
Each row corresponds to the results of the experiment where input data is identified through (i) the input data and corresponding number of active transducer elements; sparse sampling (ss), limited view (lv), virtual circle (vc), semi circle (sc), multisegment (ms) and (ii) the array type used for input; virtual circle (vc), semi circle (sc), multisegment (ms), and linear (li).
SWFD- and MSFD-mini correspond to experiment conducted on OADAT-mini dataset.}

\begin{tabular}{@{}llllllll@{}}
\toprule
                              &          & \multicolumn{2}{c}{Dice}        & \multicolumn{2}{c}{IoU}         & \multicolumn{2}{c}{HD95} \\ \midrule
                              &          & vessels        & skin curve     & vessels        & skin curve    & vessels        & skin curve  \\ \cmidrule(l){3-8} 
\textbf{SCD}                  &          &                &                &                &               &                &             \\
Full view                     & vc,vc    & 0.996 \textpm 9.8e-3 & 0.999 \textpm 8.7e-4 & 0.993 \textpm 1.8e-2 & 0.999 \textpm 1.7e-3 & 0.792 \textpm 1.0e+0 & 0.164 \textpm 2.0e+0 \\[1.5ex]
\multirow{2}{*}{Limited view} & lv128,li & 0.963 \textpm 3.4e-2 & 0.988 \textpm 7.3e-3 & 0.931 \textpm 5.5e-2 & 0.976 \textpm 1.4e-2 & 2.734 \textpm 2.9e+0 & 2.196 \textpm 7.4e+0\\
                              & lv128,vc & 0.933 \textpm 5.8e-2 & 0.978 \textpm 1.1e-2 & 0.878 \textpm 8.6e-2 & 0.957 \textpm 2.1e-2 & 4.310 \textpm 8.9e+0 & 7.372 \textpm 1.4e+1\\[1.5ex]
\multirow{6}{*}{Sparse view}  & ss128,vc & 0.980 \textpm 2.3e-2 & 0.990 \textpm 4.7e-3 & 0.961 \textpm 4.0e-2 & 0.980 \textpm 9.1e-3 & 2.499 \textpm 5.2e+0 & 3.897 \textpm 1.0e+1\\
                              & ss64,vc  & 0.967 \textpm 2.8e-2 & 0.965 \textpm 1.5e-2 & 0.937 \textpm 4.6e-2 & 0.933 \textpm 2.8e-2 & 2.738 \textpm 3.0e+0 & 9.634 \textpm 1.5e+1\\
                              & ss32,vc  & 0.938 \textpm 3.9e-2 & 0.946 \textpm 2.3e-2 & 0.886 \textpm 6.0e-2 & 0.899 \textpm 4.1e-2 & 3.074 \textpm 3.1e+0 & 12.220 \textpm 1.6e+1\\
                              & ss128,ms & 0.983 \textpm 2.4e-2 & 0.994 \textpm 2.9e-3 & 0.968 \textpm 4.1e-2 & 0.989 \textpm 5.6e-3 & 2.289 \textpm 2.4e+0 & 5.311 \textpm 1.3e+1\\
                              & ss64,ms  & 0.984 \textpm 2.4e-2 & 0.993 \textpm 3.1e-3 & 0.969 \textpm 4.1e-2 & 0.985 \textpm 6.1e-3 & 2.772 \textpm 8.2e+0 & 2.668 \textpm 8.5e+0\\
                              & ss32,ms  & 0.971 \textpm 2.8e-2 & 0.985 \textpm 4.8e-3 & 0.945 \textpm 4.5e-2 & 0.971 \textpm 9.2e-3 & 3.026 \textpm 7.0e+0 & 6.795 \textpm 1.5e+1
                              \\ \midrule
\textbf{SWFD-mini}                  &          &                &    &   &         & &      \\
\multirow{2}{*}{Full view}      & sc,sc    & 0.857 \textpm 5.7e-2 & N/A  & 0.755 \textpm 8.5e-2   &  N/A & 17.225 \textpm 2.8e+1 & N/A\\
                                & ms,ms & 0.842 \textpm 5.9e-2 &  N/A &  0.732 \textpm 8.6e-2   &  N/A& 17.218 \textpm 2.6e+1 & N/A\\[1.5ex]
\multirow{2}{*}{Limited view} & lv128,li & 0.794 \textpm 6.4e-2 &  N/A &  0.662 \textpm 8.3e-2   &  N/A& 28.685 \textpm 3.9e+1 & N/A\\
                              & lv128,sc & 0.843 \textpm 5.5e-2 &  N/A &  0.732 \textpm 7.8e-2   &  N/A& 24.444 \textpm 2.6e+1 & N/A\\[1.5ex]
\multirow{6}{*}{Sparse view}  & ss128,sc & 0.864 \textpm 4.3e-2 &  N/A &  0.763 \textpm 6.5e-2  &  N/A& 23.146 \textpm 3.0e+1 & N/A\\
                              & ss64,sc  & 0.841 \textpm 1.1e-1 &  N/A &  0.737 \textpm 1.3e-1   &  N/A& 20.216 \textpm 2.8e+1 & N/A\\
                              & ss32,sc  & 0.864 \textpm 4.1e-2 &  N/A &  0.762 \textpm 6.3e-2   &  N/A& 26.832 \textpm 3.0e+1 & N/A\\ 
                              & ss128,ms  & 0.836 \textpm 5.8e-2 &  N/A &  0.722 \textpm 8.3e-2   &  N/A& 17.909 \textpm 2.8e+1 & N/A\\
                              & ss64,ms  & 0.837 \textpm 5.6e-2 & N/A &  0.723 \textpm 8.0e-2   &  N/A& 15.811 \textpm 2.3e+1 & N/A\\ 
                              & ss32,ms  & 0.809 \textpm 6.0e-2 &  N/A &  0.684 \textpm 8.2e-2   & N/A& 20.603 \textpm 3.1e+1 & N/A \\ \midrule
\textbf{MSFD-mini}                 &          &                         &      &   &     & &    \\
Limited view                     & lv128,li & 0.474 \textpm 1.3e-1 &     N/A     &  0.320 \textpm 1.2e-1 &  N/A& 20.134 \textpm 3.3e+1 & N/A\\[1.5ex]
\multirow{3}{*}{Sparse view}  & ss128,ms & 0.563 \textpm 1.0e-1 &    N/A    & 0.400 \textpm 1.1e-1  &  N/A& 14.312 \textpm 1.9e+1 & N/A\\
                              & ss64,ms  & 0.572 \textpm 1.2e-1 &    N/A    & 0.411 \textpm 1.3e-1  &  N/A& 19.716 \textpm 2.1e+1 & N/A\\
                              & ss32,ms  & 0.639 \textpm 1.4e-1 &     N/A    & 0.485 \textpm 1.5e-1  &  N/A& 11.499 \textpm 1.7e+1 & N/A\\
                              \bottomrule
\end{tabular}
\label{tab:segmentation}
\end{table}

\section{Discussion}
Major differences exist between simulated and experimental datasets.
Even if the content is different, training and test samples of the simulated dataset are inherently sampled from the same distribution.
On the other hand, experimental datasets feature shifts due to different volunteers being imaged, inherent noise from data acquisition system, and difference in directional sensitivity resulting from transducer alignment and positioning of the hand-held probe.
Furthermore, despite the efforts to avoid corrupted acquisitions during the data collection, experimental datasets still contain samples with relatively low signal-to-noise ratio.
Such samples are expected to yield reduced performance metrics for image translation tasks due to significant mismatch between the predicted and noisy target images.
In a clinical setting, a medical expert typically repeats an acquisition if they deem the signal quality is significantly lower than expected. 
However, beyond this subjective filtering step, one needs to make sure that even the worst results are either sufficiently good or their poor performance can be attributed to a cause. 
Accordingly, we further analyze some of the worst samples in the Appendix and believe that this should be a standard for future work.
Provided dataset is limited to one body part of volunteers without any known health issues.
The image reconstruction methods for OA imaging can be applied on any other body part or imaging setup as they solve the same physical inverse problem.
The image translation algorithms for sparse acquisition and limited view problems can be adapted for different devices and acquisitions by another clinician/technician at another center, as the streak artifacts originating from sparse acquisition and limited view follow the same pattern.

We envision that future research will tackle additional challenges such as unsupervised or weakly supervised domain adaptation across the datasets provided in this work.
There are initial studies to correct limited view artifacts in OA using transfer learning between simulated and experimental datasets after domain adaptation~\citep{klimovskaia2022signal}.
Similarly, transfer learning between simulated and experimental domains can enhance segmentation performance of the vessels and skin curve in clinical images.
Using properties of the detected tissues for fluence correction and heterogeneous SoS image reconstructions would then yield more accurate and quantitative images~\citep{deanben2014effects}.
We anticipate additional contributions in the field of representation learning using OADAT, e.g.\,through self-supervised learning, could allow overcoming bottlenecks for specialized downstream tasks with limited amount of task-specific available data.
The multispectral datasets with paired images across multiple wavelength acquisitions are expected to facilitate the investigation of generative modeling of multispectral signals from a given wavelength.
We also anticipate future work to explore novel multispectral unmixing approaches using MSFD, enabling more accurate quantification of oxygenation, melanin and lipid content of the tissues.
Finally, the provided raw signal data, not available in commercial devices, has a high value for data-driven research.
For example, it can serve to benchmark methodologies in image reconstruction e.g., based on variational networks with loop unrolling~\citep{vishnevskiy2019deep}, as well as ultra-fast imaging through adaptive channel sampling.

\section{Conclusion}

In this work, we provide experimental and synthetic clinical OA data covering a large variety of examples to be used as common ground to compare established and new data processing methods.
The datasets correspond to samples from volunteers of varying skin types, different OA array geometries, and several illumination wavelengths.
A subset of this experimental data is annotated by an expert.
The dataset is supplemented with simulated samples containing ground truth acoustic pressure maps, annotations, and combine pairs of samples reconstructed with different OA array geometries.
We define a set of 44 experiments tackling major challenges in the OA field and provide reconstructions of the images under these scenarios along with their corresponding ground truths. 
We propose and release 44 \footnote{Pretrained model weights and various scripts to train and evaluate modUNet are available at \href{https://renkulab.io/gitlab/firat.ozdemir/oadat-evaluate}{https://renkulab.io/gitlab/firat.ozdemir/oadat-evaluate}.}
trained neural networks that achieve a good performance for all these examples which can be used as baselines for future improvements.
Additional problems can further be defined with the data provided, such as the effects of random sparse sampling or the presence of noise in the signal matrices prior to reconstruction.
We believe that these datasets and benchmarks will play a significant role in fostering coordinated efforts to solve major challenges in OA imaging.

\section*{Broader Impact Statement}

The dataset is anonymized by randomly assigning identity numbers for each volunteer.
The true identities will not be accessible by any third party now or in the future. 
The dataset should not be used to draw any medical conclusions.
The purpose of the dataset is to help researchers to develop new image processing methods and provide benchmark scores.

\section*{Author Contributions}
B.L. setup and acquired raw datasets, B.L. and F.O. analyzed and parsed datasets, F.O. conducted numerical experiments based on the defined tasks. All authors reviewed the manuscript.

\section*{Acknowledgments}
This work was supported by Swiss Data Science Center (grant C19-04).

\medskip
\bibliography{main}

\newpage

\appendix
\section{Optoacoustic imaging background}
\label{sec:optoacoustic_imaging_background}
Optoacoustic (OA) imaging setups contain light source, transducer array, data acquisition system and workstation PC.
Usually, nanosecond-duration pulsed lasers are used as light sources.
Alternatively, light emitting diodes (LEDs) can be used with high repetition rates (shorter pulses) as cheaper alternatives to lasers.
The excitation of objects/tissues with short, pulsed light sources results in heating and expansion (thermoelastic expansion) of the materials.
This expansion generates OA signals (US waves) that propagate to various directions in a 3D imaging medium (see \ref{sec:optoacoustic_equation} for detailed explanation of pressure waves).
Transducer arrays are used to capture the propagating waves in the imaging medium.
Piezocomposite elements in these arrays convert mechanical pressure waves into electrical signals (see \ref{sec:transducer_array_details} for transducer array specifications).
After the detection and conversion, the electrical signals are digitized by the data acquisition systems at the defined sampling rate.
The signals are acquired and digitized by all the elements in a transducer array simultaneously.
The OA acquisition generates time domain (raw) signals acquired from each transducer element.
Image reconstruction is performed by solving the inverse problem of signal to image domain conversion.
In this study, we used backprojection method to reconstruct OA images (see \ref{sec:oa_image_reconstruction} for the detailed explanation of the method and used algorithms).
Overall data acquisition system and experimental design are explained in \ref{sec:experimental_data_acquisition_setup}.

\subsection{Optoacoustic wave equations}
\label{sec:optoacoustic_equation}
OA imaging is based on thermoelastic expansion of the tissues which results in propagation of pressure waves in imaging medium depending on spatial and temporal changes. The OA wave equation can be written as follows
\begin{center}
$\frac{\partial^2 p(r,t)}{\partial t^2} - c^2 \nabla^2p(r,t) = \Gamma H(r,t)\frac{\partial\delta (t)}{\partial t}$
\end{center}
where $r$ and $t$ are the spatial and temporal variables, respectively.
$\Gamma$ is the Gr\"uneisen constant \citep{gruneisen1912theorie}.
$c$ stands for speed of sound.
$H(r,t)$ is the absorbed energy field based on the location and time of the sample.
$\delta (t)$ stands for temporal laser light intensity change based on the illumination.
$p(r,t)$ represents the pressure wave dependent on spatial and temporal variables.
The Poisson solution of OA wave equation for pressure wave can be written as
\begin{center}
$ p(r,t) = \frac{\Gamma}{4\pi c} \frac{\partial}{\partial t} \int_{S'} \frac{H(r')}{|r-r'|} dS'$
\end{center}
where $S'$ is the time dependent spherical surface defined by $|r-r'|=ct$.
This equation represents OA forward model which inverse problem of reconstruction can be derived.
OA images are reconstructed by absorbed energy field $H(r')$ at specific location based on measured pressure waves.
$H(r')$ is calculated from detected pressure waves at the surface S as follows
\begin{center}
$ H(r') = \frac{1}{\Gamma} \int_{\Omega} \frac{d\Omega}{\Omega} \left[2p(r,t)-2t\frac{\partial p(r,t)}{\partial t}\right]_{t=\frac{r-r'}{c}}$.
\end{center}
The constants at this equation can be omitted.
After omitting the constants in the formula, the equation is discretized as  
\begin{center}
$ H(r'_{j}) = \sum \left[p(r_{i}, t_{ij}) - t_{ij}\frac{\partial p(r_{i},t_{ij})}{\partial t} \right] $
\end{center}
where $r'_{j}$ is the $j$-th point on the defined reconstruction grid, $r_{i}$ is the position of $i$-th transducer and $t_{ij} = |r_{i}-r'_{j}|/c$.
In summary, the equation calculates the distance between a point on the defined grid and an element of the transducer array.
Then, it finds the corresponding wave intensity in signal (or defined as surface) based on the time of flight calculated using speed of sound in the imaging medium.

\subsection{Transducer array details}
\label{sec:transducer_array_details}
\textbf{\textit{Semi circle:}} The array contains 256  piezocomposite transducers distributed over a semi circle (concave surface) equidistantly with the radius of 40 mm (Fig.~\ref{fig:experimental_data}b, main manuscript).
The single transducer elements have dimensions of 0.37 mm × 15 mm with inter-element distance of 0.10 mm.
This configuration of transducer elements results in cylindrical (toroidal) focusing at 38 mm (close to the center of the array).
The central peak frequency of array is 5 MHz with 60\% bandwidth at -6dB.
\\
\textbf{\textit{Multisegment:}} The array is formed by the combination of a linear detector array and concave parts on the right and left sides as shown in Fig.~\ref{fig:experimental_data}c (main manuscript).
The linear part contains 128 elements distributed on a linear surface with inter-element pitch size of 0.25 mm.
Both of the concave parts include 64 elements which make the total number of elements equal to 256 (128 linear + 128 concave).
The inter-element pitch size of concave part is 0.6 mm with 40 mm radius of curvature.
The height of all elements are equal to 10 mm.
Concave parts are designed to increase angular coverage in OA imaging.
This configuration results in a cylindrical focusing at 38 mm close to the center of the array.
The array has 7.5 MHz central frequency with 70\% bandwidth at -6 dB.
\\
\textbf{\textit{Linear array:}} The array is central part of the multisegment array with 128 transducer elements distributed over a line with pitch size of 0.25 mm (Fig.~\ref{fig:experimental_data}c, main manuscript).
Similar to concave parts, the linear array has 7.5 MHz central frequency with 70\% bandwidth at -6 dB.
The linear array is optimized for US data acquisitions with planar waves.
Hence, the array produces OA images with limited view artifacts due to reduced angular coverage which is a limiting factor for OA image acquisitions.
\\
\textbf{\textit{Virtual circle:}} The array is generated to simulate images with 360 degree angular coverage which results in artifact free reconstructions (Fig.~\ref{fig:simulated_data}a, main manuscript).
It contains 1,024 transducer elements distributed over a full circle with equal distance.
The radius of the transducer array is kept equal to semi circle array (40 mm) to allow comparison between simulations and experimental acquisitions.

\subsection{OA image reconstruction}
\label{sec:oa_image_reconstruction}
A Python package ``{\href{https://github.com/berkanlafci/pyoat}{pyoat}'' is presented with the datasets to reconstruct images from raw signals.
The library uses ``pip'' package manager to install and use the functions.
The package provides functions to bandpass filter and normalize raw signals as preprocessing step.
Implementation of backprojection algorithm is also included in the package (Sec. \ref{sec:reconstruction_methods}, main manuscript).
The forward model operator is implemented to simulate signals from acoustic pressure maps.
Data readers and savers are integrated into the package for loading the raw data and saving the reconstructed images, respectively.
The package provides examples to use different functions.
The examples are easy to use Python scripts which require only raw signal data paths as input.
Positions of elements in all transducer arrays used in this study are included in the library to enable reconstruction from raw signals.

\subsection{Experimental data acquisition setup}
\label{sec:experimental_data_acquisition_setup}
Signal acquisition is performed with OA imaging setup that combines four main components, namely, transducer arrays, nanosecond pulsed lasers, data acquisition system and workstation PC (Fig. \ref{fig:experimental_data}, main manuscript).
SWFD dataset was acquired with multisegment and semi circle transducer arrays (Sec. \ref{sec:transducer_arrays}, main manuscript) using a nanosecond laser at 1,064 nm with repetition rate of 10 Hz.
MSFD dataset was acquired with only multisegment array at six different wavelengths (700, 730, 760, 780, 800, 850 nm) using the laser with repetition rate of 50 Hz.
The increased repetition rate guarantees that displacement within the scene between the frames of different wavelengths are minimal.
Data acquisition system is used to digitize signals acquired by the transducer arrays.
A sampling rate of 40 mega-samples-per-second was used in all experiments.
Then, digital signals are sent to workstation PC to store the data and display both raw signals and reconstructed images in real time with lower resolution.
The high resolution images presented in this work were reconstructed after the acquisition (offline) using the backprojection method (Sec. \ref{sec:reconstruction_methods}, main manuscript).
The real time feedback helps to position the transducer arrays.
The workstation PC also synchronizes all the imaging setup components by setting delays for data acquisition system and triggers for laser pulses. 
Imaging medium was filled with water to increase coupling efficiency between the transducer arrays and the skin.
Water has low attenuation at the wavelengths used in this study.
Hence, the signal attenuation in the medium was low enough to be neglected.
Transducer arrays are held orthogonal to the forearm surface throughout acquisition and swiftly moved from elbow towards wrist.
All participants joined the experiments voluntarily and were informed about details of the experiments.

\newpage

\section{Fitzpatrick skin phototype in experimental datasets}
\label{sec:fitzpatrick_skin_phototype_in_experimental_datasets}
\begin{wrapfigure}{r}{0.40\textwidth}
\vspace{-1.5em}
  \begin{center}
    \begin{subfigure}[b]{0.499\linewidth}
    \centering
        \includegraphics[width=\textwidth, trim= 0cm 0.25cm 0cm 0cm, clip]{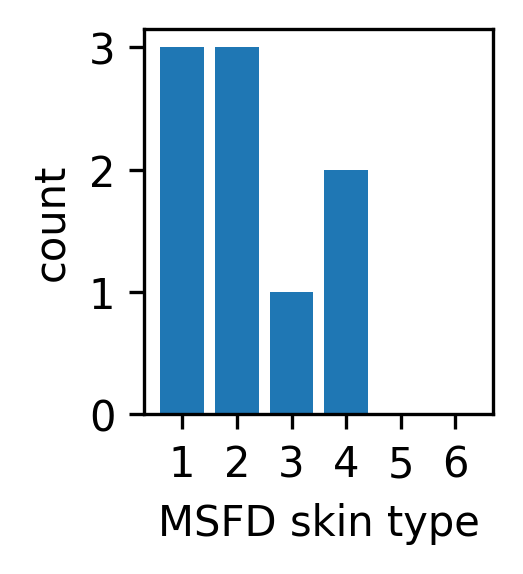}
    \end{subfigure}\hfill
    \begin{subfigure}[b]{0.499\linewidth}
    \centering
        \includegraphics[width=\textwidth, trim= 0cm 0.25cm 0cm 0cm, clip]{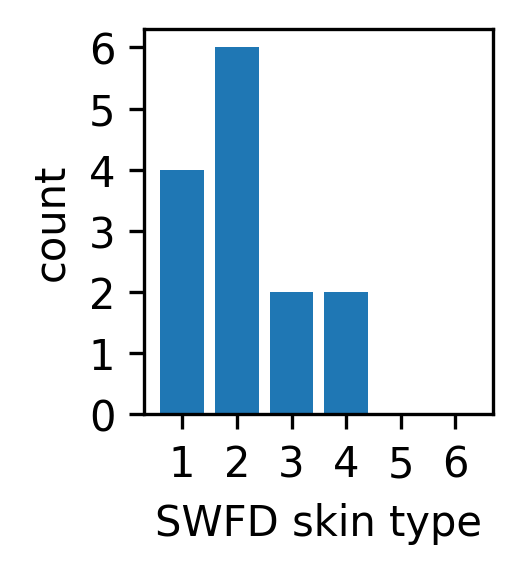}
    \end{subfigure}
  \end{center}
  \caption{Fitzpatrick skin phototype~\citep{gupta2019skin} distribution of volunteers in datasets.}
    \label{fig:skin_type_counts}
\end{wrapfigure}

Fitzpatrick skin phototype is a metric to quantify the amount of melanin pigment in the skin of a subject \citep{gupta2019skin}. 
The metric ranges from one to six, going from pale white skin to black skin color.
This is relevant for various OA applications due to different skin types (melanin concentration) lead to varying amount of contrast (absorption) at skin surfaces in the acquired images.
Accordingly, the distribution of the skin types of the volunteers across SWFD and MSFD are shown in Fig.~\ref{fig:skin_type_counts}.

\section{Architecture and implementation details}
\label{sec:architecture_and_implementation_details}

\begin{figure}
  \begin{center}
  \centering
  \includegraphics[width=0.99\textwidth]{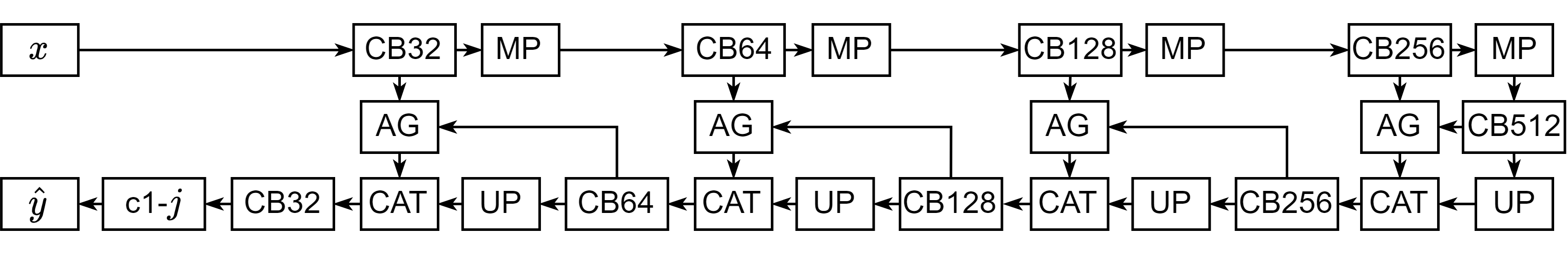}
  \end{center}
  \caption{Schematic of the proposed modUNet architecture. 
CB$j$ represents the residual 2D convolutional block with batch normalization shown in Fig.~\ref{fig:resconvblockBN} (main manuscript) where each convolution has $j$ filters.
Other abbreviations correspond to 2D-maxpooling (MP) of poolsize 2, 2D bilinear upsampling (UP) by a factor of 2, concatenation (CAT), and attention gates (AG)~\citep{oktay2018attention}.
Finally, c1-$j$ represents a convolutional layer of $j$ filters (1 for image translation and 3 for semantic segmentation experiments) and $1 \times 1$ kernels without activation. }
  \label{fig:modified_unet_schematic}
\end{figure}

We show the schematic of the proposed modUNet architecture in Fig.~\ref{fig:modified_unet_schematic}.

We use categorical cross entropy loss and mean squared error loss for segmentation and image translation experiments, respectively. 
We add an additional $l_1=l_2=0.01$ regularization weight for each learned model parameter.

For all experiments, we use Adam optimizer and scale the learning rate with exponential decay (decay rate of 0.98 and decay steps of 1,000) from a peak of $0.0001$ following a linear warmup of 10,000 steps. 
Except for experiments on OADAT-mini, all models are trained for 150 epochs with a mini-batch size of 25 and best validation set loss is used as the early stopping criteria for which test set performance metrics are presented. 
Experiments done on OADAT-mini are optimized for 10,000 epochs.
For each experiment, we used an Nvidia P100 or Titan X GPU with 12GB memory to train the model. 
We do not augment data during training. 
Experiments on MSFD, SWFD, and SCD took about, 61, 32 and 15 hours, respectively. 
The training time variation across datasets is due to the difference in size of the datasets. 

\section{Qualitative results}
\label{sec:qualitative_results}
Despite the impressive performance of the modUNet, there is a long tail in almost all taks for SSIM or IoU distributions as seen in Figs.~\ref{fig:image_translation} and~\ref{fig:semantic_segmentation} (main manuscript). 
Accordingly, we qualitatively examine some of the worst samples at the bottom of these tails for each task.

\subsection{Image translation}

We showcase and analyze a selection of the worst performing samples from the most challenging limited view and sparse sampling reconstruction experiments on SWFD and report SSIM for these samples with respect to the target sample. 

In Figs.~\ref{fig:qualitative_image_translation_swfd_lv128li}~\&~\ref{fig:qualitative_image_translation_swfd_lv128sc} (experiments $f_{\mathrm{SWFD\_lv128, li}}$ and $f_{\mathrm{SWFD\_lv128, sc}}$) one can see that some of} the worst samples based on SSIM correspond to images with little to no signal content. 
The remaining sample from the 10th percentile in Fig.~\ref{fig:qualitative_image_translation_swfd_lv128li} suggests that low SSIM in multisegment array images can be deceiving due to different set of artifacts that appear on the target (multisegment) array.
In Fig.~\ref{fig:qualitative_image_translation_swfd_lv128sc}, where the target is the semi circle array, the artifacts are not as intense. 
Nevertheless, given the accurate vessel geometry correction, low SSIMs in 1st, 5th and 10th percentiles can be attributed to other local noise patterns, such as the circular noise that exist on the target image.
Thanks to the majority of samples in our datasets having high SNR, most of the undesirable artifacts that seldomly occur in test samples are removed.
For example, in Figs.~\ref{fig:qualitative_image_translation_swfd_lv128sc}~\&~\ref{fig:qualitative_image_translation_msfd_lv128li}, columns with relatively higher signal content are either very close to the target or even more artifact-free.

In Figs.~\ref{fig:qualitative_image_translation_swfd_ss32sc}~\&~\ref{fig:qualitative_image_translation_swfd_ss32ms}, we show some of the samples with the worst SSIM from the experiments $f_{\mathrm{SWFD\_ss32, sc}}$ and $f_{\mathrm{SWFD\_ss32, ms}}$, respectively. 
The samples in the left two columns for both experiments have very poor SNR.
For all other samples, modUNet predictions from the heavily subsampled image reconstructions indicate higher SNR than the target samples thanks to removing most of the noise pattern (especially visible within the water medium prior to/above the forearm) while keeping all soft tissue structures intact and faithful to their counterparts in the target image.
Unfortunately, mismatch of the noise patterns causes low SSIM for such samples with low SNR target images.

For completeness, we show the worst performing samples for SSIM in image translation experiments using SCD in Fig.~\ref{fig:qualitative_image_translation_scd_ss32vc_scd_lv128li}.
One can see that the rounded SSIM already reaches 1.0 by the 1st- (right figure) or 5th- (left figure) percentile performing sample in the test set. 
In the left figure 2nd column, one can see that the strong artifact passing through the small vessel caused modUNet to separate the vessel into two pieces, shown with a red arrow.

In addition, we also showcase some of the best performing samples for SSIM in Figs.~\ref{fig:qualitative_image_translation_swfd_lv128sc_best},~\ref{fig:qualitative_image_translation_swfd_lv128li_best},~\ref{fig:qualitative_image_translation_msfd_lv128li_best},~\ref{fig:qualitative_image_translation_swfd_ss32sc_best}~\&~\ref{fig:qualitative_image_translation_swfd_ss32ms_best}.
Beyond correcting distorted vessel geometry, it can be seen in some samples that certain vessels that are barely visible get accurately redrawn to match the target (e.g., Fig.~\ref{fig:qualitative_image_translation_swfd_lv128sc} 3rd and 4th columns, Fig.~\ref{fig:qualitative_image_translation_swfd_ss32sc} 2nd, 3rd and 4th column, Fig.~\ref{fig:qualitative_image_translation_swfd_ss32ms} 3rd and 4th columns, and Fig.~\ref{fig:qualitative_image_translation_swfd_ss32ms_best} 4th column).\\

\begin{figure}
\centering
\includegraphics[width=0.73\textwidth, trim= 0cm 0.0cm 0cm 0cm, clip]{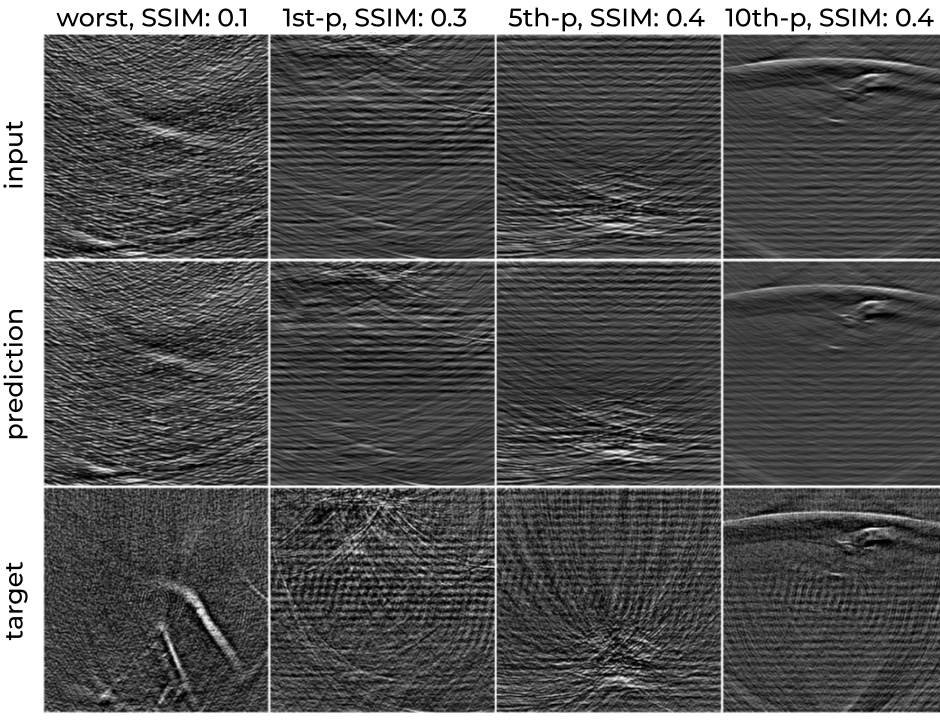}
\caption{We showcase the worst (1st column),1st-} (2nd column), 5th- (3rd column), and 10th-percentile (4th column) SSIM samples based on modUNet predictions for experiment $f_{\mathrm{SWFD\_lv128, li}}$ with input (1st row), modUNet prediction (2nd row), and target (3rd sample) pairs.
Red arrows indicate some of the distorted vessel geometries at input getting corrected at modUNet predictions.
\label{fig:qualitative_image_translation_swfd_lv128li}
\end{figure}

\begin{figure}
\centering
\includegraphics[width=0.73\textwidth, trim= 0cm 0.0cm 0cm 0cm, clip]{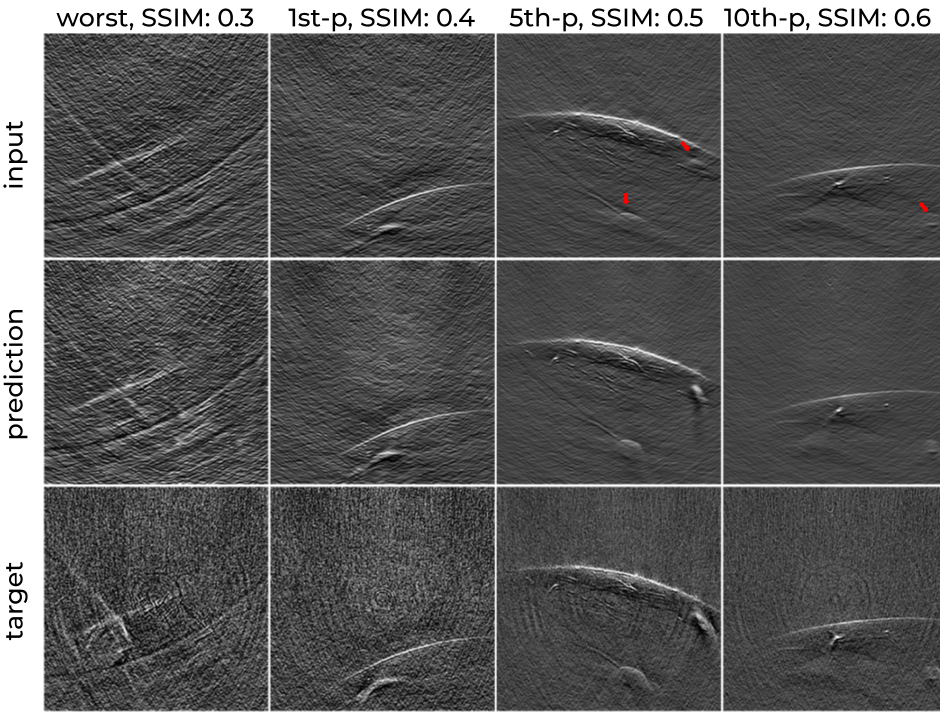}
\caption{We showcase the worst (1st column), 1st-} (2nd column), 5th- (3rd column), and 10th-percentile (4th column) SSIM samples based on modUNet predictions for experiment $f_{\mathrm{SWFD\_lv128, sc}}$ with input (1st row), modUNet prediction (2nd row), and target (3rd sample) pairs.
Red arrows indicate some of the distorted vessel geometries at input getting corrected at modUNet predictions.
\label{fig:qualitative_image_translation_swfd_lv128sc}
\end{figure}

\begin{figure}[]
\centering
\includegraphics[width=0.73\textwidth, trim= 0cm 0.0cm 0cm 0cm, clip]{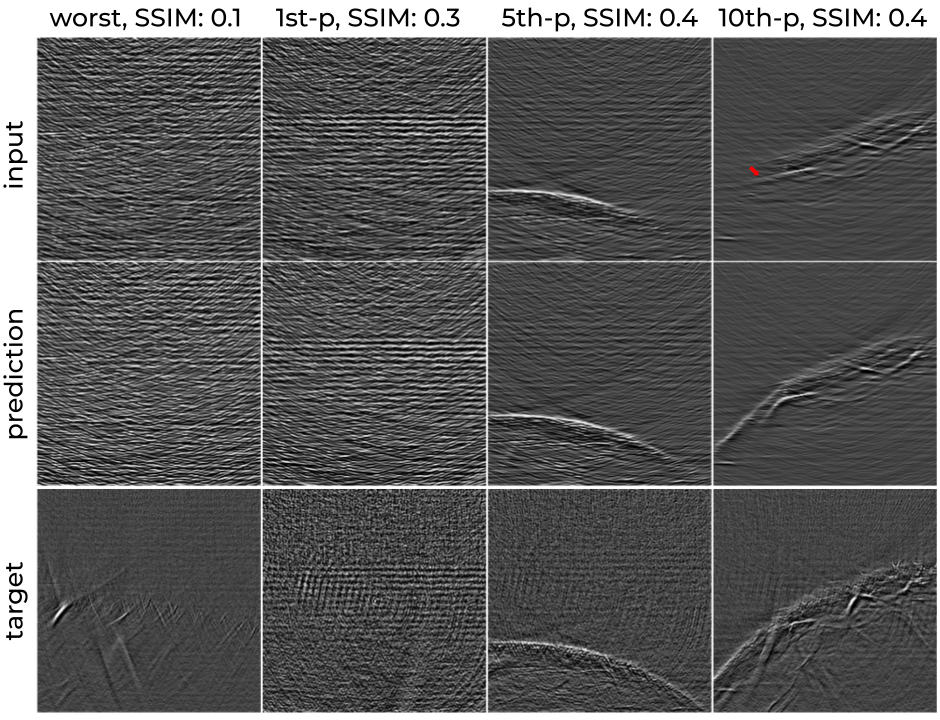}
\caption{We showcase the worst (1st column), 1st-} (2nd column), 5th- (3rd column), and 10th-percentile (4th column) SSIM samples based on modUNet predictions for experiment $f_{\mathrm{MSFD\_lv128, li}}$ with input (1st row), modUNet prediction (2nd row), and target (3rd sample) pairs.
Red arrows indicate some of the distorted vessel geometries at input getting corrected at modUNet predictions.
\label{fig:qualitative_image_translation_msfd_lv128li}
\end{figure}

\begin{figure}[]
\centering
\includegraphics[width=0.73\textwidth, trim= 0cm 0.0cm 0cm 0cm, clip]{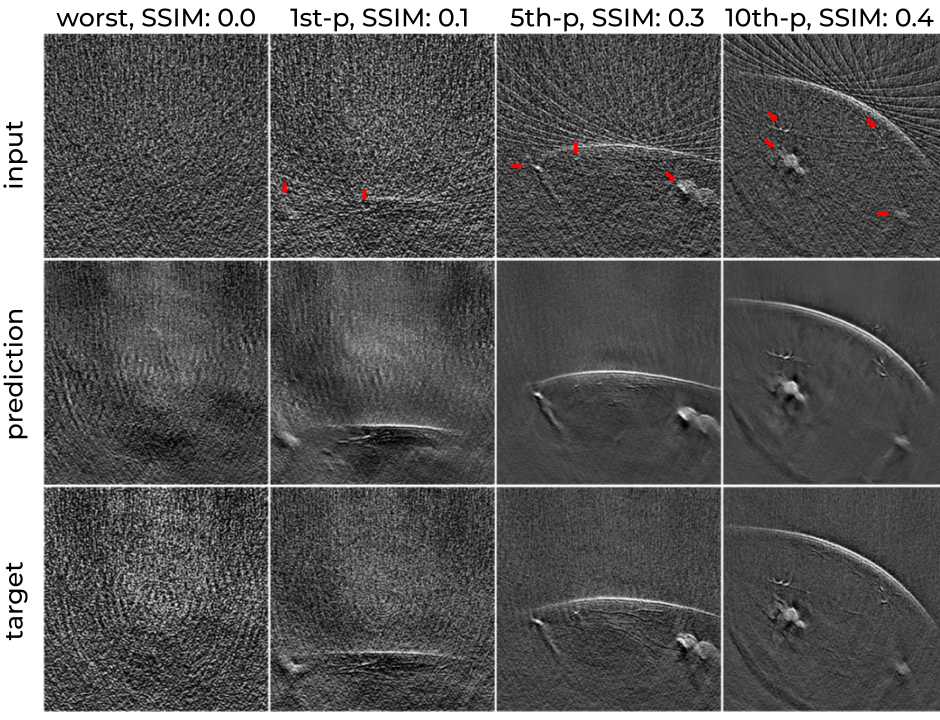}
\caption{We showcase the worst (1st column), 1st-} (2nd column), 5th- (3rd column), and 10th-percentile (4th column) SSIM samples based on modUNet predictions for experiment $f_{\mathrm{SWFD\_ss32, sc}}$ with input (1st row), modUNet prediction (2nd row), and target (3rd sample) pairs.
\label{fig:qualitative_image_translation_swfd_ss32sc}
\end{figure}

\begin{figure}
\centering
\includegraphics[width=0.73\textwidth, trim= 0cm 0.0cm 0cm 0cm, clip]{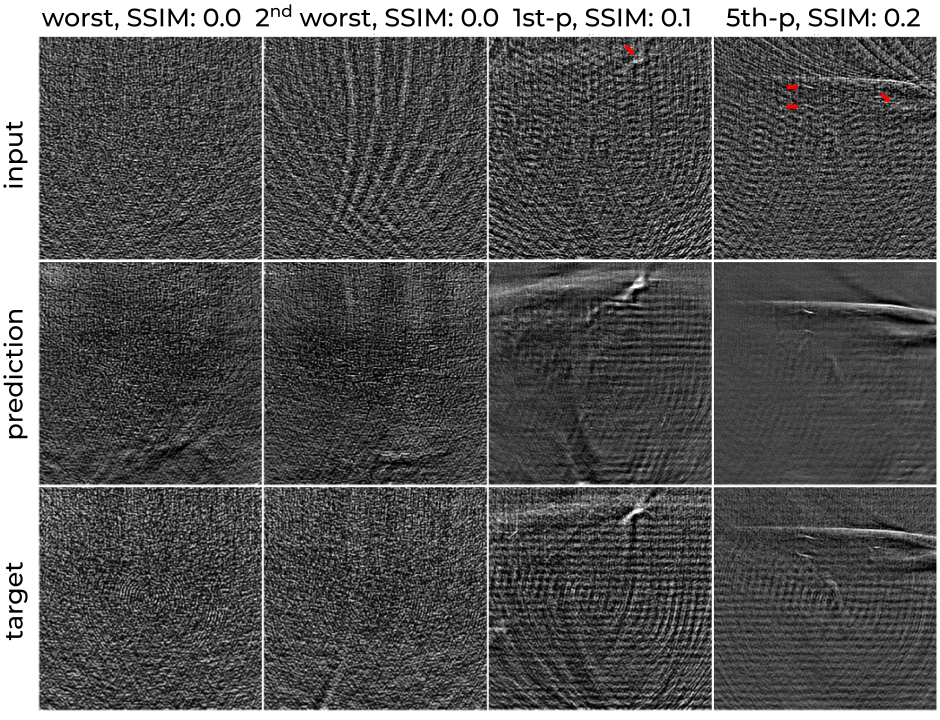}
\caption{
We showcase the worst (1st column), 1st- (2nd column), 5th- (3rd column), and 10th-percentile (4th column) SSIM samples based on modUNet predictions for experiment $f_{\mathrm{SWFD\_ss32, ms}}$ with input (1st row), modUNet prediction (2nd row), and target (3rd sample) pairs.
Red arrows indicate some of the nearly invisible vessels at input getting corrected at modUNet predictions.}

\label{fig:qualitative_image_translation_swfd_ss32ms}
\end{figure}

\begin{figure}
\centering
\begin{subfigure}[b]{0.495\linewidth}
\centering
\includegraphics[width=\textwidth, trim= 0cm 0.0cm 0cm 0cm, clip]{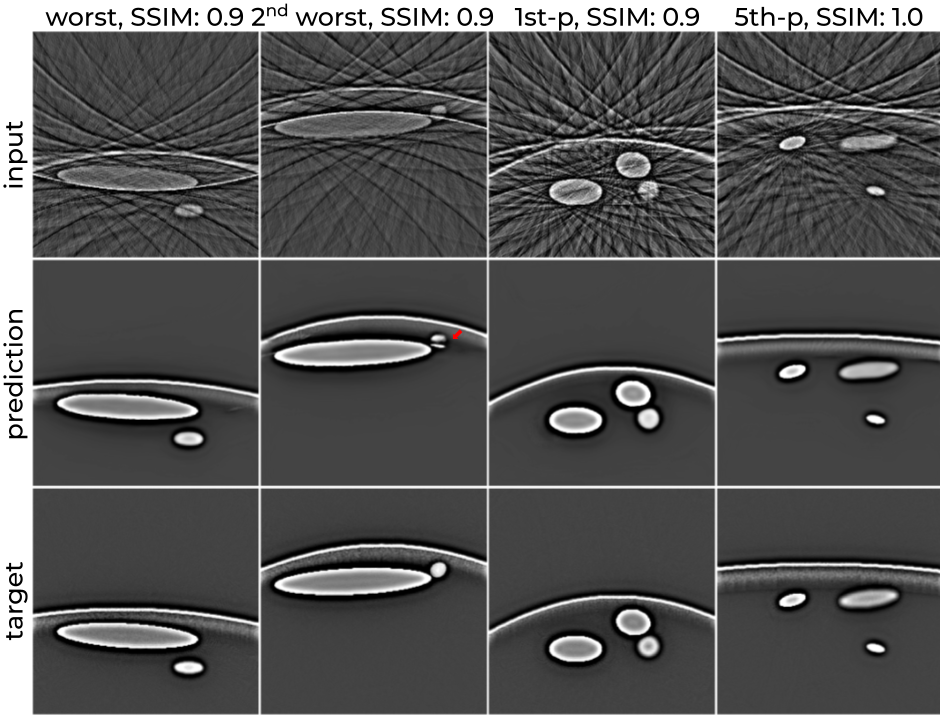}
\end{subfigure}
\begin{subfigure}[b]{0.495\linewidth}
\centering
\includegraphics[width=\textwidth, trim= 0cm 0.0cm 0cm 0cm, clip]{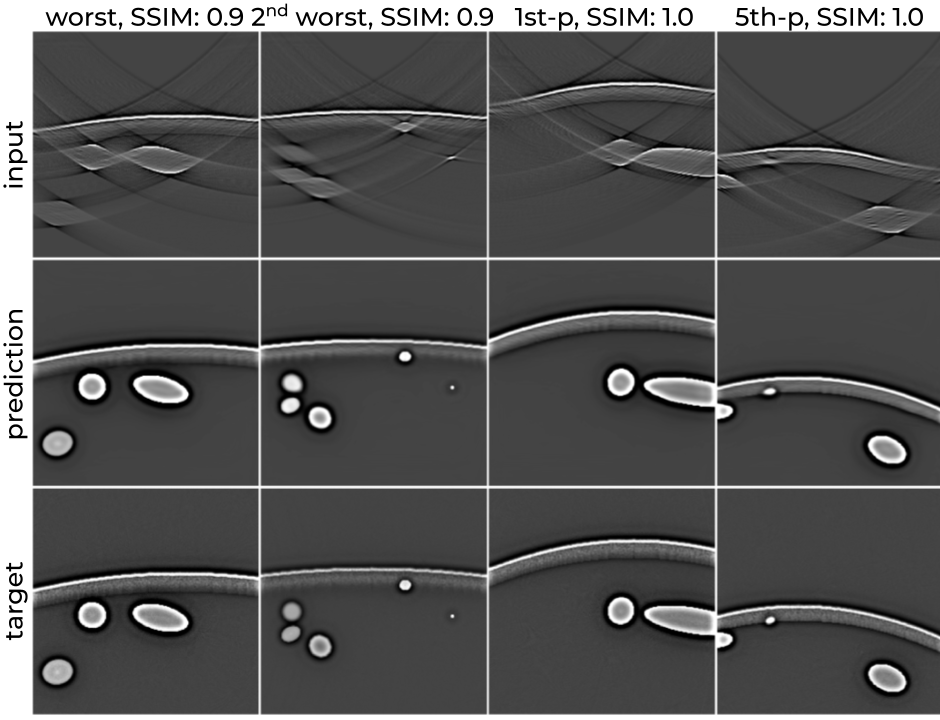}
\end{subfigure}\hfill
\caption{We showcase the worst (1st column), 2nd worst (2nd column), 1st- (3rd column), and 5th-percentile (4th column) SSIM samples based on modUNet predictions for experiments $f_{\mathrm{SCD\_ss32, vc}}$ (left) and $f_{\mathrm{SCD\_lv128, li}}$ (right) with input (1st row), modUNet prediction (2nd row), and target (3rd sample) pairs.
}
\label{fig:qualitative_image_translation_scd_ss32vc_scd_lv128li}
\end{figure}

\begin{figure}
\centering
\includegraphics[width=0.73\textwidth, trim= 0cm 0.0cm 0cm 0cm, clip]{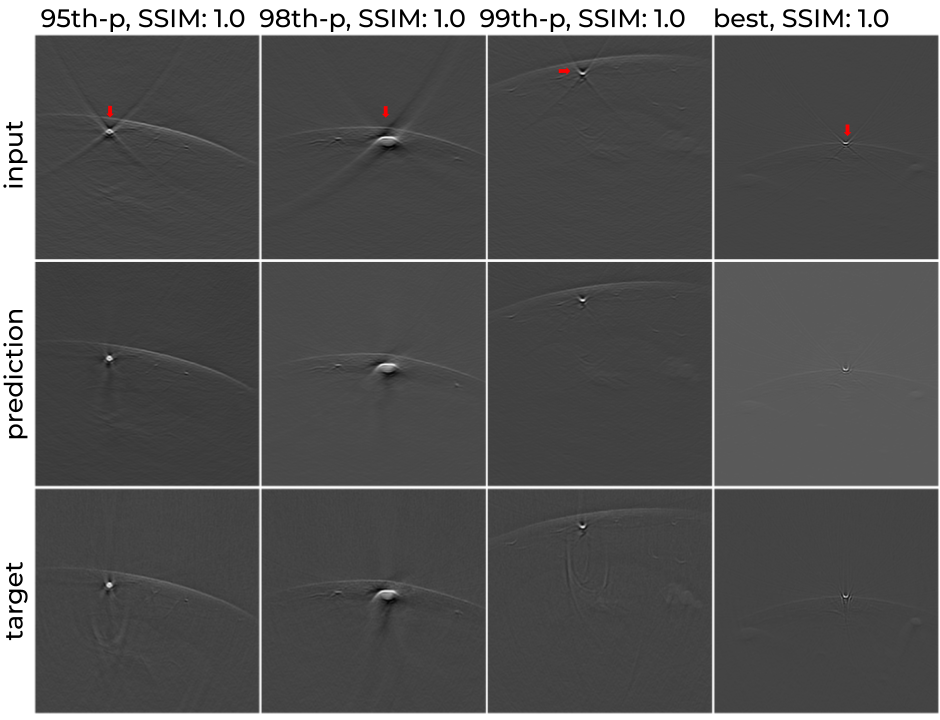}
\caption{We showcase 95th- (1st column), 98th- (2nd column), 99th-percentile (3rd column), and the best (4th column) SSIM samples based on modUNet predictions for experiment $f_{\mathrm{SWFD\_lv128, sc}}$ with input (1st row), modUNet prediction (2nd row), and target (3rd sample) pairs.
}
\label{fig:qualitative_image_translation_swfd_lv128sc_best}
\end{figure}

\begin{figure}
\centering
\includegraphics[width=0.73\textwidth, trim= 0cm 0.0cm 0cm 0cm, clip]{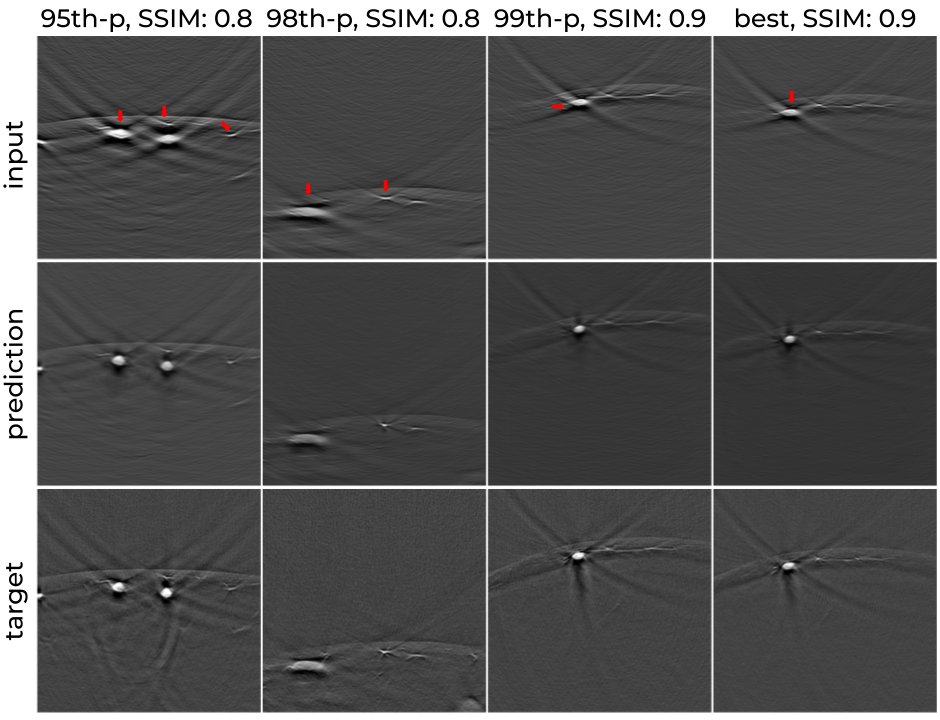}
\caption{We showcase 95th- (1st column), 98th- (2nd column), 99th-percentile (3rd column), and the best (4th column) SSIM samples based on modUNet predictions for experiment $f_{\mathrm{SWFD\_lv128, li}}$ with input (1st row), modUNet prediction (2nd row), and target (3rd sample) pairs.
Red arrows indicate some of the distorted vessel geometries at input getting corrected at modUNet predictions.}
\label{fig:qualitative_image_translation_swfd_lv128li_best}
\end{figure}

\begin{figure}
\centering
\includegraphics[width=0.73\textwidth, trim= 0cm 0.0cm 0cm 0cm, clip]{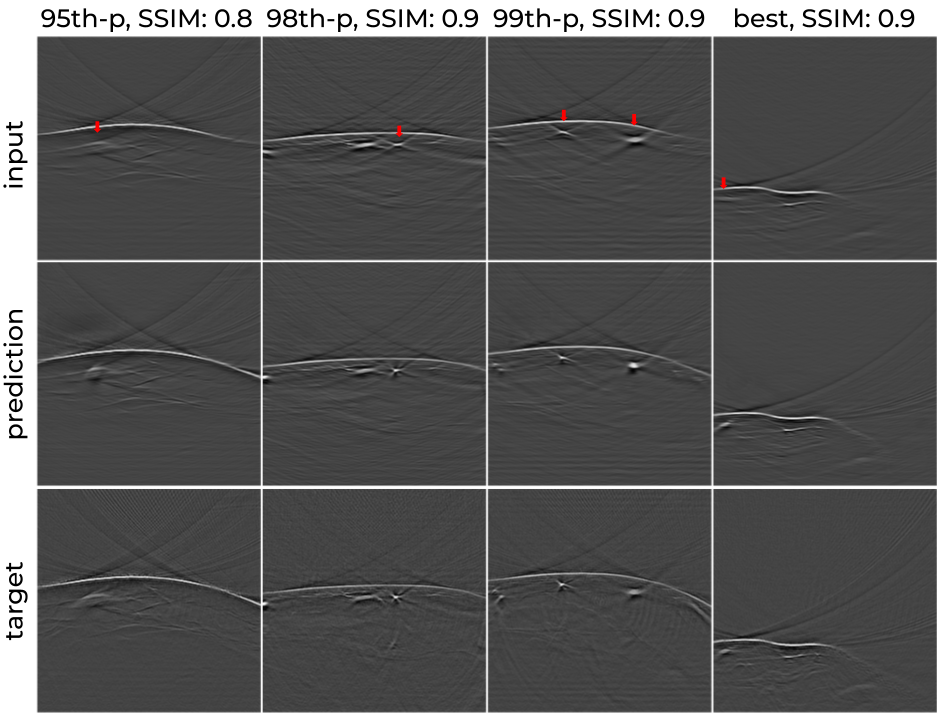}
\caption{We showcase 95th- (1st column), 98th- (2nd column), 99th-percentile (3rd column), and the best (4th column) SSIM samples based on modUNet predictions for experiment $f_{\mathrm{MSFD\_lv128, li}}$ with input (1st row), modUNet prediction (2nd row), and target (3rd sample) pairs.
Red arrows indicate some of the distorted vessel geometries at input getting corrected at modUNet predictions.}
\label{fig:qualitative_image_translation_msfd_lv128li_best}
\end{figure}

\begin{figure}
\centering
\includegraphics[width=0.73\textwidth, trim= 0cm 0.0cm 0cm 0cm, clip]{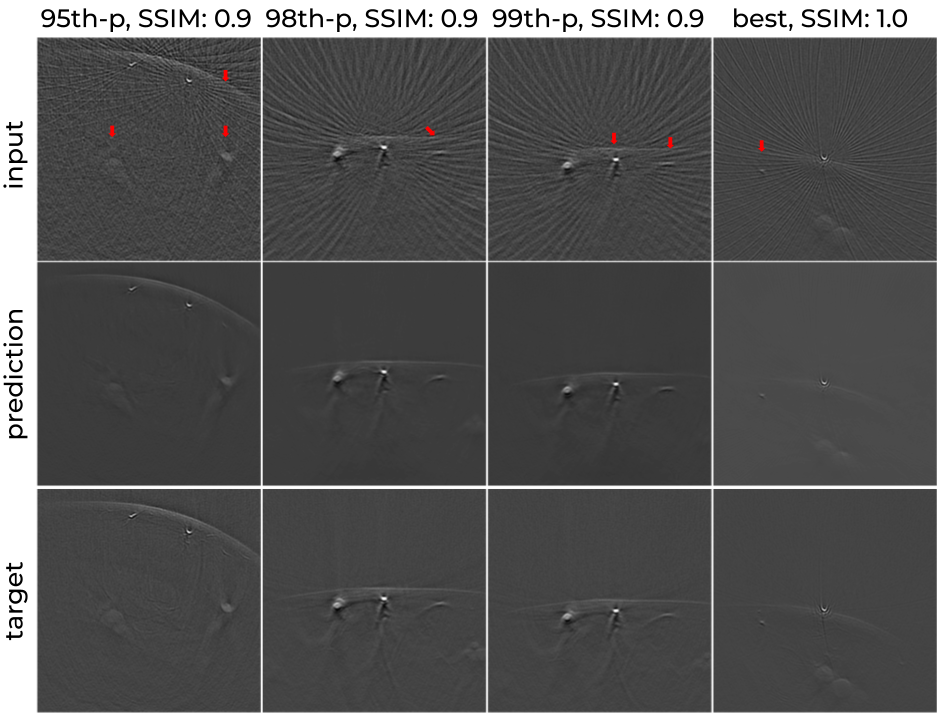}
\caption{We showcase 95th- (1st column), 98th- (2nd column), 99th-percentile (3rd column), and the best (4th column) SSIM samples based on modUNet predictions for experiment $f_{\mathrm{SWFD\_ss32, sc}}$ with input (1st row), modUNet prediction (2nd row), and target (3rd sample) pairs.
}
\label{fig:qualitative_image_translation_swfd_ss32sc_best}
\end{figure}

\begin{figure}[!t]
\centering
\includegraphics[width=0.73\textwidth, trim= 0cm 0.0cm 0cm 0cm, clip]{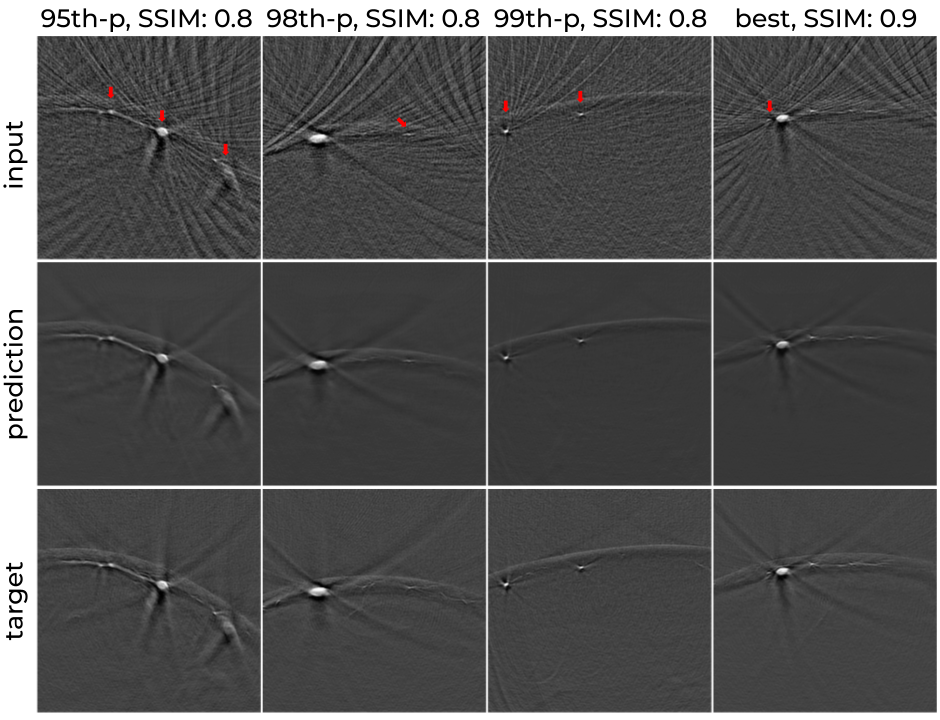}
\caption{
We showcase 95th- (1st column), 98th- (2nd column), 99th-percentile (3rd column), and the best (4th column) SSIM samples based on modUNet predictions for experiment $f_{\mathrm{SWFD\_ss32, ms}}$ with input (1st row), modUNet prediction (2nd row), and target (3rd sample) pairs.}

\label{fig:qualitative_image_translation_swfd_ss32ms_best}
\end{figure}

\subsection{Semantic segmentation}

\begin{figure}[!hb]
\centering
\includegraphics[width=0.73\textwidth, trim= 0cm 0.0cm 0cm 0cm, clip]{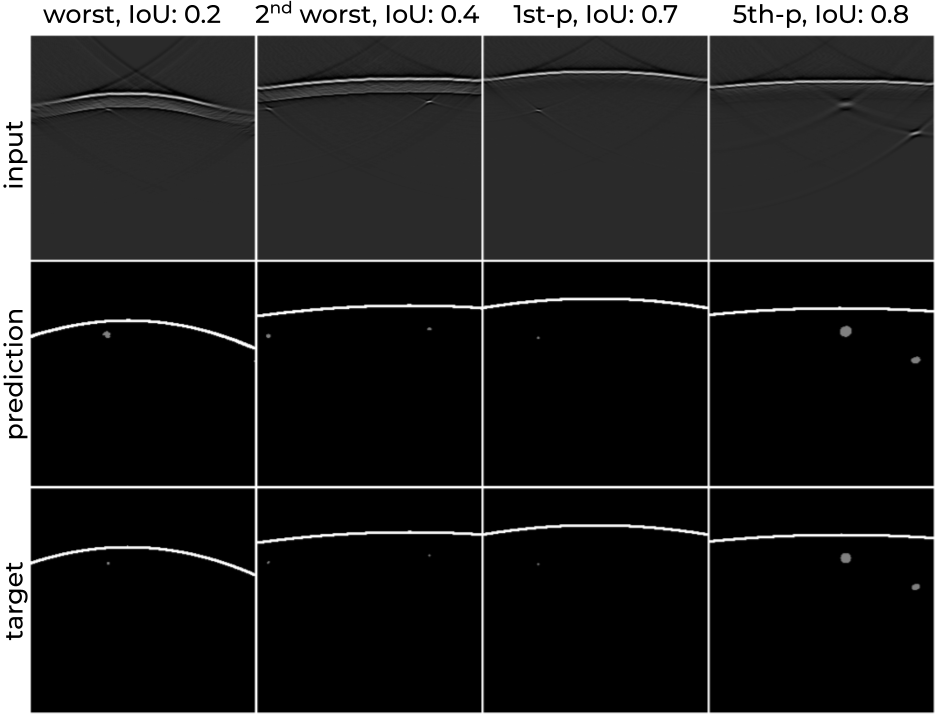}
\caption{We showcase the worst (1st column), 2nd worst (2nd column), 1st- (3rd column), and 5th-percentile (4th column) vessel IoU samples based on modUNet predictions for experiment $f_{\mathrm{seg\_lv128, li}}$ with input (1st row), modUNet prediction (2nd row), and ground truth (3rd sample) pairs.}
\label{fig:qualitative_semantic_segmentation_seg_lv128li}
\end{figure}

\begin{figure}
\centering
\includegraphics[width=0.73\textwidth, trim= 0cm 0.0cm 0cm 0cm, clip]{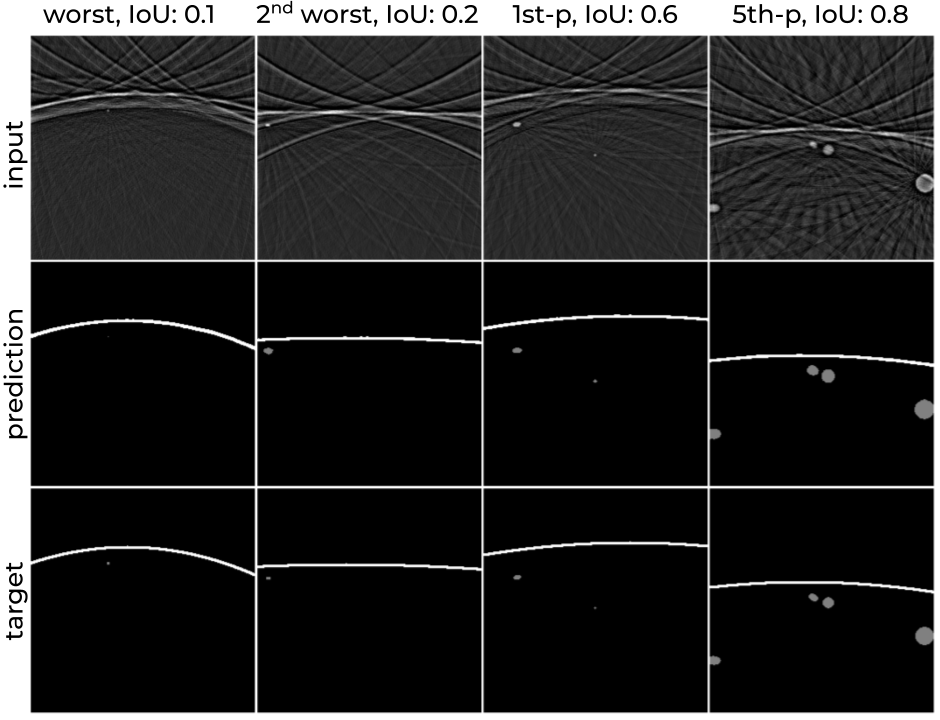}
\caption{We showcase the worst (1st column), 2nd worst (2nd column), 1st- (3rd column), and 5th-percentile (4th column) vessel IoU samples based on modUNet predictions for experiment $f_{\mathrm{seg\_ss32, vc}}$ with input (1st row), modUNet prediction (2nd row), and ground truth (3rd sample) pairs.}
\label{fig:qualitative_semantic_segmentation_seg_ss32vc}
\end{figure}

\begin{figure}
\centering
\includegraphics[width=0.73\textwidth, trim= 0cm 0.0cm 0cm 0cm, clip]{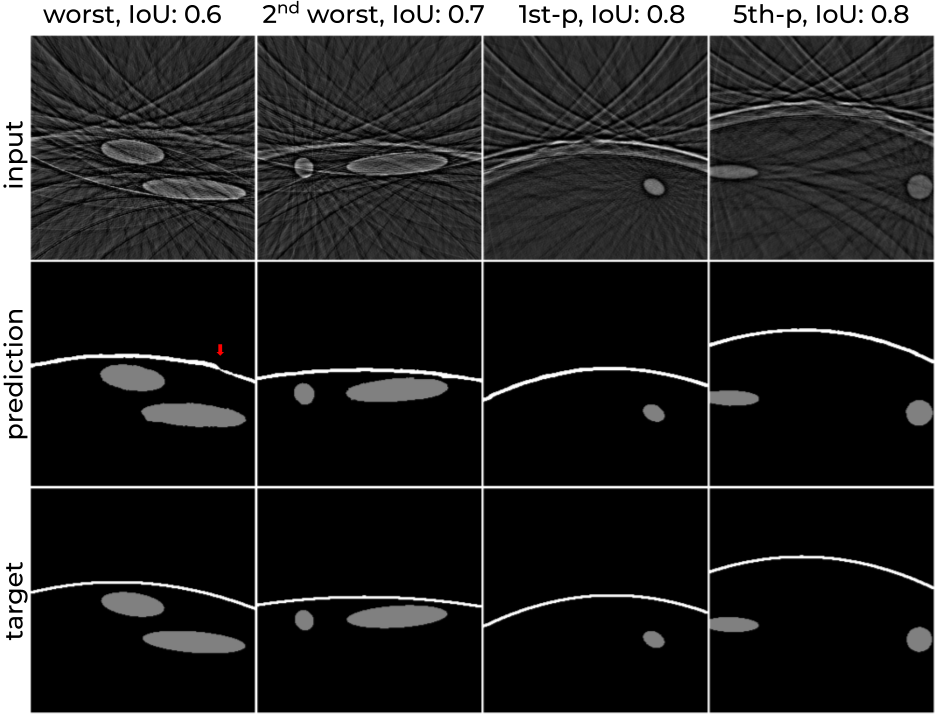}
\caption{We showcase the worst (1st column), 2nd worst (2nd column), 1st- (3rd column), and 5th-percentile (4th column) skin curve IoU samples based on modUNet predictions for experiment $f_{\mathrm{seg\_ss32, vc}}$ with input (1st row), modUNet prediction (2nd row), and ground truth (3rd sample) pairs.}
\label{fig:qualitative_semantic_segmentation_seg_ss32vc_skin}
\end{figure}

\begin{figure}[!t]
\centering
\includegraphics[width=0.73\textwidth, trim= 0cm 0.0cm 0cm 0cm, clip]{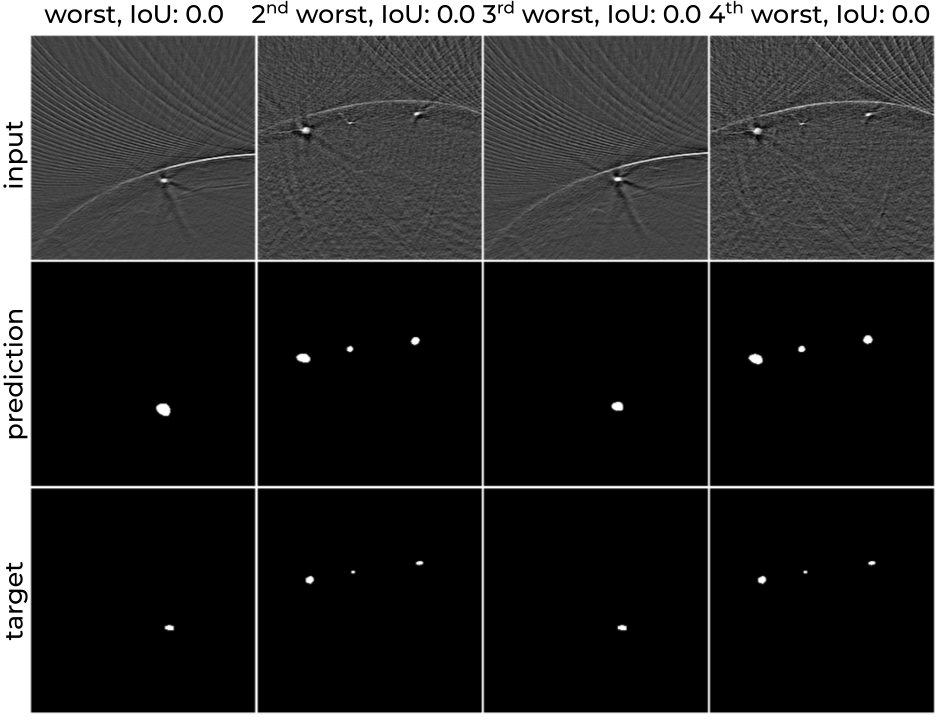}
\caption{We showcase the worst (1st column), 2nd- (2nd column), 3rd- (3rd column), and 4th-worst (4th column) IoU samples based on modUNet predictions for experiment $f_{\mathrm{seg\_MSFD\_ss64, ms}}$ with input (1st row), modUNet prediction (2nd row), and target (3rd sample) pairs.}%
\label{fig:qualitative_semantic_segmentation_seg_msfd_ss64ms}
\end{figure}

\begin{figure}
\centering
\begin{subfigure}[b]{0.495\linewidth}
\centering
\includegraphics[width=\textwidth, trim= 0cm 0.0cm 0cm 0cm, clip]{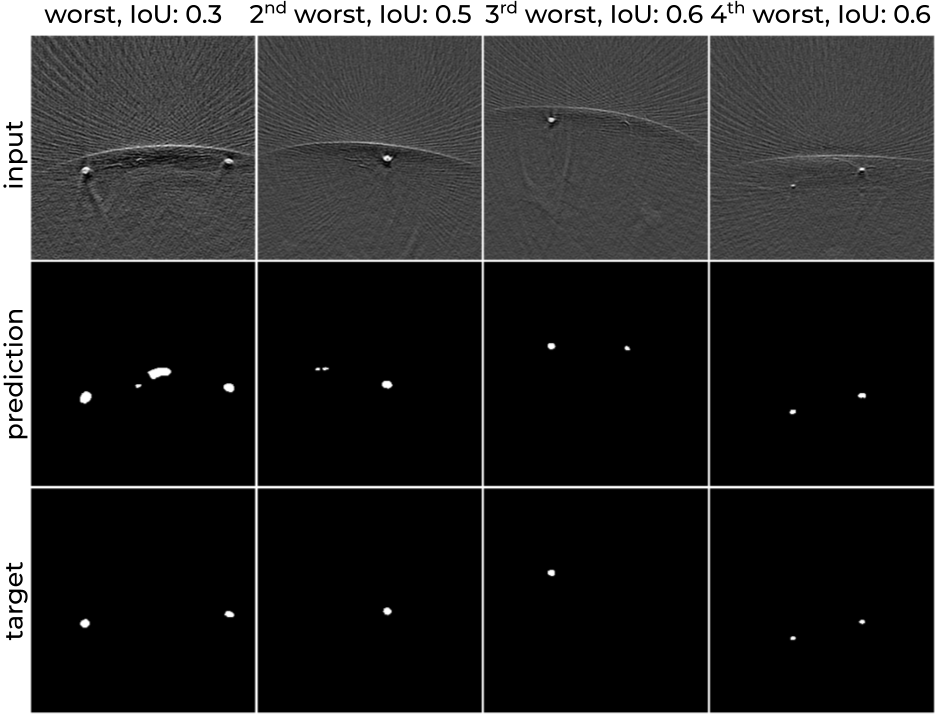}
\end{subfigure}
\begin{subfigure}[b]{0.495\linewidth}
\centering
\includegraphics[width=\textwidth, trim= 0cm 0.0cm 0cm 0cm, clip]{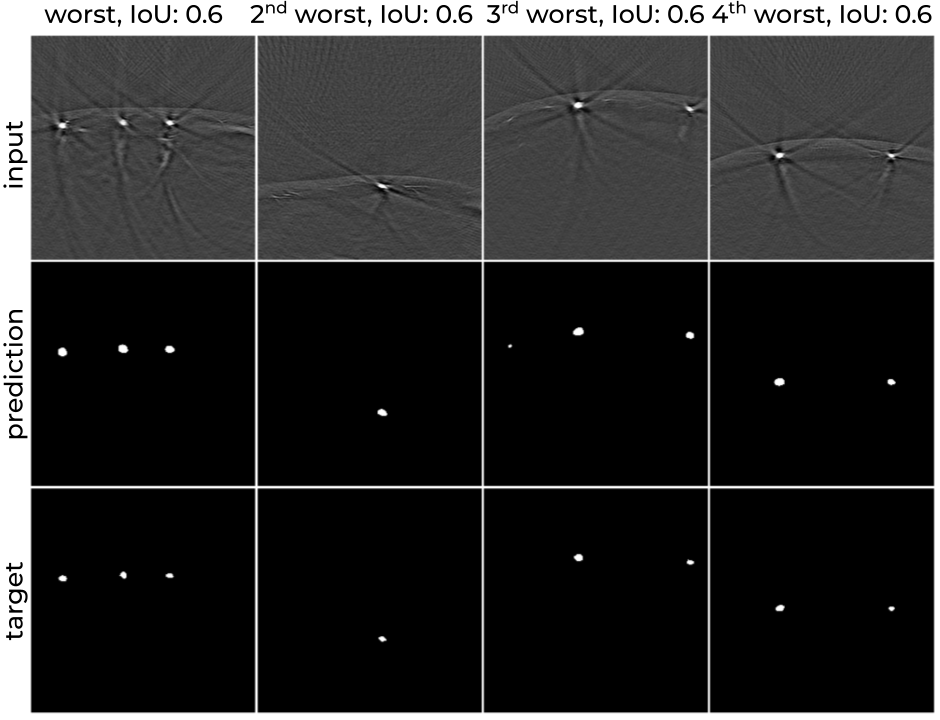}
\end{subfigure}\hfill
\caption{We showcase the worst (1st column), 2nd- (2nd column), 3rd- (3rd column), and 4th-worst (4th column) IoU samples based on modUNet predictions for experiments $f_{\mathrm{seg\_SWFD\_ss64,sc}}$ (left) and $f_{\mathrm{seg\_SWFD\_ss128, ms}}$ (right) with input (1st row), modUNet prediction (2nd row), and target (3rd sample) pairs.}
\label{fig:qualitative_semantic_segmentation_seg_swfd_ss64sc_swfd_ss128ms}
\end{figure}

Similar to image translation task, we look at some of the samples from SCD with the worst vessel segmentation performance with respect to IoU metric in Figs.~\ref{fig:qualitative_semantic_segmentation_seg_lv128li}~\&~\ref{fig:qualitative_semantic_segmentation_seg_ss32vc} (experiments $f_{\mathrm{seg\_lv128, li}}$ and $f_{\mathrm{seg\_ss32, vc}}$).
As a result of SCD having flawless ground truth annotations, modUNet can achieve near perfect segmentation performance for all classes in all experiments of both tasks. 
The decrease of IoU seems to be exclusive for under- or over-segmenting little vessels by a handful of pixels. 
Despite most vessels being small (see Fig.~\ref{fig:simulated_data}e in the main manuscript, distribution of number of pixels per vessel), 
low IoU is limited to only a handful of extreme cases. 
This is also corroborated by the fact that skin curve segmentation distribution (see Fig.~\ref{fig:semantic_segmentation} right in the main manuscript) having a significantly shorter tail, barely falling below an IoU of 0.95 in most experiments. 
In order to understand the single experiment with a considerable performance drop, we look at experiment $f_{\mathrm{seg\_ss32, vc}}$ in Fig.~\ref{fig:qualitative_semantic_segmentation_seg_ss32vc_skin}.
It can be observed that under heavy sparse sampling, the skin curve often gets ambiguous, rarely leading to suboptimal segmentation performance in these cases.
Nevertheless, for most such subsampled reconstructions having intense duplicates of skin curve (Fig.~\ref{fig:qualitative_semantic_segmentation_seg_ss32vc} first row), modUNet successfully segments the right curve.  
Considering OADAT-mini segmentation experiments, in Fig.~\ref{fig:qualitative_semantic_segmentation_seg_msfd_ss64ms}, we show the worst performing four samples from the test set of MSFD-mini. 
In Fig~\ref{fig:qualitative_semantic_segmentation_seg_swfd_ss64sc_swfd_ss128ms}, it can be seen that such poor IoU is not observed in the worst performing samples for SWFD-mini experiments.
\\

\newpage
\section{Simulated cylinder dataset generation}
\label{sec:simulated_cylinder_dataset_generation}

Our proposed simulated dataset, SCD, follows a group of heuristics we derived from observations on experimental images. 
For any given sample, we first generate an acoustic pressure map, from which we also generate ground truth annotations.
We then apply certain post processing steps to imitate other phenomena such as patterns under the skin surface and different vessel textures. 
Given the geometry of the transducers we want to simulate, we then apply forward transform that gives the raw signals.
Finally, we reconstruct these signals using backprojection algorithm to generate the images used in various SCD experiments in this manuscript.

The acoustic pressure map generation consists of initially drawing the curve that represents the laser pulse absorption on the skin surface mainly due to melanin.
Given that experimental data is acquired with making sure that forearm is roughly at a certain distance range from the arrays, we also limit the drawn skin curve distance.
We define the skin curve as a 2nd degree polynomial that is fitted to three points randomly sampled at the two horizontal edges and the center of the image at varying heights. 
As a post-processing step to mimic experimental data, the curve is first smoothed with a Gaussian filter.
Then, an exponential decay of randomized length is applied under the curve along vertical axis. 
Finally, a non-structured uniform normal noise is multiplied with the aforementioned exponential decay region.
For vessel generation, the number of cylinders to be drawn is sampled based on a coin flip.
Based on the outcome, either two cylinders drawn or the number of cylinders is sampled from Poisson distribution (see Fig.~\ref{fig:simulated_data}e, distribution of number of vessels per image, main manuscript).
Each vessel is initially represented by a cylinder orthogonal to the image plane (z-axis) with a randomly sampled radius. 
We then randomly rotate the cylinder around x- and y- axes. 
The vessel is determined as the cross-section of the cylinder at the imaging plane, yielding ellipses based on the final angle of the cylinders.
As a post-processing step, we flip a coin to determine whether the vessel has a homogeneous intensity profile or has a linearly decreasing intensity from its center.
We then apply a Gaussian filter on the vessel to smooth its edges.
Finally, based on a coin flip, we decide whether or not to multiply the intensity profile of the vessel with uniform normal noise.
The same process is iteratively repeated until desired number of non-overlapping vessels are generated.
All parameters used for the aforementioned steps are empirically selected based on our observations on the experimental datasets.
We provide the script to simulate acoustic pressure maps that we used for SCD. 

\section{Organization of datasets}
\begin{figure}[!b]
\centering
\includegraphics[width=1\textwidth]{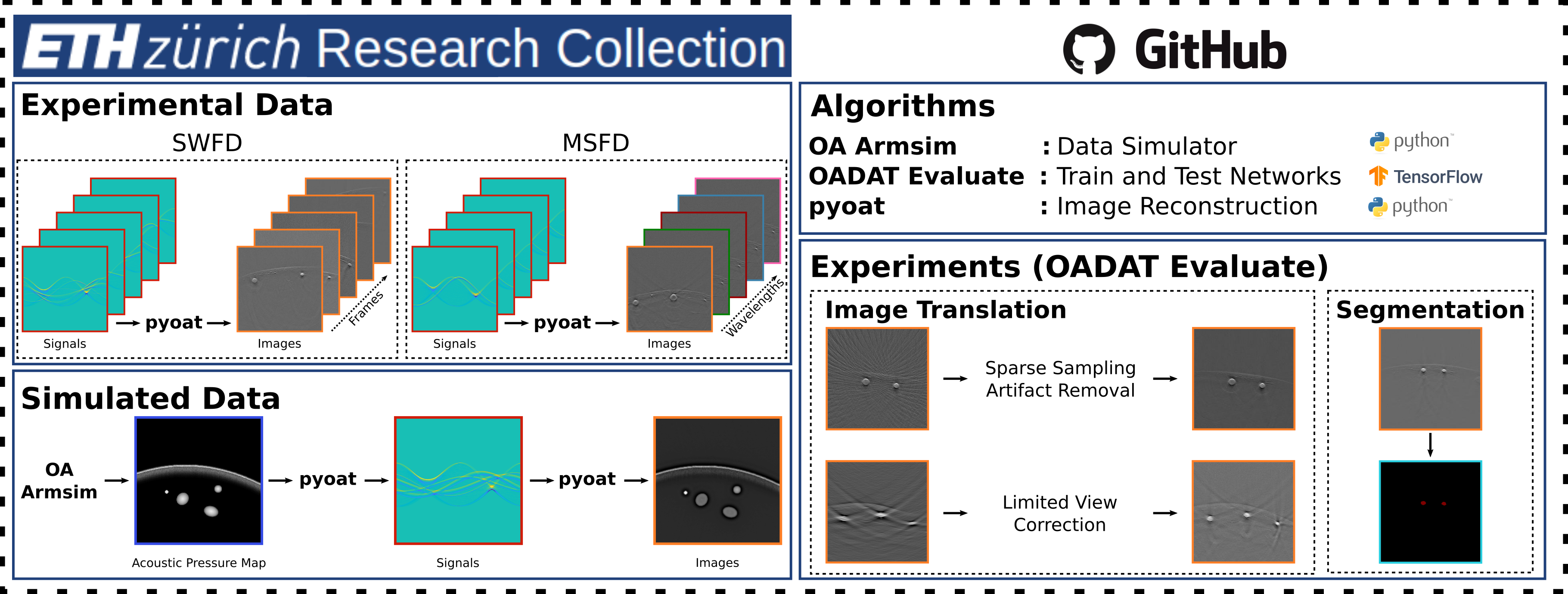}
\caption{
Pipeline figure to summarize data storage and algorithms.}
\label{fig:pipeline_figure}
\end{figure}
The datasets and required algorithms to read raw signals and reconstructed images are summarized in Fig.~\ref{fig:pipeline_figure} along with the codes to rerun and evaluate all the presented experiments.
All datasets are stored as HDF5 files and corresponding indices across datasets within an HDF5 file have the same scene as raw signal or reconstructed image.
Similarly, metadata such as \textit{patientID}, \textit{side} (i.e., left vs right arm), \textit{skin type}, and \textit{sliceID} contain the information for the corresponding indices. 
Some of these metadata such as \textit{patientID}, which corresponds to the unique anonymized identifier of the volunteer, only exists in experimental datasets. 
Matching identifiers across experimental datasets (MSFD and SWFD) correspond to the same volunteer.

Each sample (raw signal or image reconstruction) is chunked in a single piece for optimal compression/decompression overhead when reading individual images and storing datasets.
For example, for a given HDF5 dataset of shape (number of instances, height, width), each block of (1, height, width) is compressed individually.
In our experiments, this ensured zero additional idle GPU time reading random training sample from a compressed HDF5 file when compared with its non-compressed HDF5 counterpart.

Raw signal datasets have the shape of (number of instances, temporal acquisition axis, receiving element axis). 
Temporal acquisition axis is fixed to 2,030 sampling points across all transducer arrays. 
Receiving element axis depends on the number of transducer elements that is used for signal acquisition. 
For sparse sampling and limited view reconstructions, we retain the shape of the raw signal data, however, we switch off the receiving elements that are not active, hence they are padded with zeros. 
This convention also allows for direct compatibility with the \textit{pyoat} reconstruction package.

We provide a header dataset file ``OADAT.h5'' as a convenient access point to each other HDF5 dataset file.
For convenience, we also provide a header dataset file ``OADAT\_v2.h5'' that allows more intuitive access to addendums of MSFD, SWFD, and SCD with multisegment sparse sampling datasets. 
Additionally, a header dataset file ``OADAT-mini.h5'' is provided as an access point to reach all relevant OADAT-mini HDF5 files.
These header files also contain metadata module of the Dataset Nutrition Label, which should help making the dataset self explanatory.

\subsection{MSFD}

Contents of MSFD are listed in Table~\ref{tab:msfd_contents}.
The contents of the dataset are stored in the files ``MSFD\_multisegment\_RawBP.h5'' and ``MSFD\_multisegment\_ss\_RawBP.h5''.
Provided \textit{sliceID} corresponds to the time index of the slice acquired from a given volunteer for a given acquisition. 
For example, \textit{sliceID} $i+1$ is recorded right after $i$ for a given \textit{patientID} and \textit{side}.
This information can be relevant for future work that does not treat each slice independently, but exploit correlations from consecutive slice acquisitions.
Nevertheless, a \textit{sliceID} $i \in [1,..., 1400]$ does not necessarily correspond to the same position on the forearm across the volunteers.

\begin{table}[]
\caption{Contents of the MSFD file.}
\tiny
\centering
\begin{tabular}{@{}lllllll@{}}
\toprule
Name               & Shape              & Content                           &  & Name                 & Shape              & Content               \\ \cmidrule(r){1-3} \cmidrule(l){5-7} 
patientID          & (25200,)           & $(2,5,6,7,9,10,11,14,15)$         &  & ms\_ss64\_BP\_w700   & (25200, 256, 256)  & image reconstructions \\
side               & (25200,)           & $(\mathrm{left}, \mathrm{right})$ &  & ms\_ss64\_BP\_w730   & (25200, 256, 256)  & image reconstructions \\
skin\_type         & (25200,)           & $(1,2,3,4)$                       &  & ms\_ss64\_BP\_w760   & (25200, 256, 256)  & image reconstructions \\
sliceID            & (25200,)           & $(1,...,1400)$                    &  & ms\_ss64\_BP\_w780   & (25200, 256, 256)  & image reconstructions \\
linear\_BP\_w700   & (25200, 256, 256)  & image reconstructions             &  & ms\_ss64\_BP\_w800   & (25200, 256, 256)  & image reconstructions \\
linear\_BP\_w730   & (25200, 256, 256)  & image reconstructions             &  & ms\_ss64\_BP\_w850   & (25200, 256, 256)  & image reconstructions \\
linear\_BP\_w760   & (25200, 256, 256)  & image reconstructions             &  & ms\_ss128\_BP\_w700  & (25200, 256, 256)  & image reconstructions \\
linear\_BP\_w780   & (25200, 256, 256)  & image reconstructions             &  & ms\_ss128\_BP\_w730  & (25200, 256, 256)  & image reconstructions \\
linear\_BP\_w800   & (25200, 256, 256)  & image reconstructions             &  & ms\_ss128\_BP\_w760  & (25200, 256, 256)  & image reconstructions \\
linear\_BP\_w850   & (25200, 256, 256)  & image reconstructions             &  & ms\_ss128\_BP\_w780  & (25200, 256, 256)  & image reconstructions \\
ms\_BP\_w700       & (25200, 256, 256)  & image reconstructions             &  & ms\_ss128\_BP\_w800  & (25200, 256, 256)  & image reconstructions \\
ms\_BP\_w730       & (25200, 256, 256)  & image reconstructions             &  & ms\_ss128\_BP\_w850  & (25200, 256, 256)  & image reconstructions \\
ms\_BP\_w760       & (25200, 256, 256)  & image reconstructions             &  & ms\_ss32\_raw\_w700  & (25200, 2030, 256) & raw signals           \\
ms\_BP\_w780       & (25200, 256, 256)  & image reconstructions             &  & ms\_ss32\_raw\_w730  & (25200, 2030, 256) & raw signals           \\
ms\_BP\_w800       & (25200, 256, 256)  & image reconstructions             &  & ms\_ss32\_raw\_w760  & (25200, 2030, 256) & raw signals           \\
ms\_BP\_w850       & (25200, 256, 256)  & image reconstructions             &  & ms\_ss32\_raw\_w780  & (25200, 2030, 256) & raw signals           \\
linear\_raw\_w700  & (25200, 2030, 256) & raw signals                       &  & ms\_ss32\_raw\_w800  & (25200, 2030, 256) & raw signals           \\
linear\_raw\_w730  & (25200, 2030, 256) & raw signals                       &  & ms\_ss32\_raw\_w850  & (25200, 2030, 256) & raw signals           \\
linear\_raw\_w760  & (25200, 2030, 256) & raw signals                       &  & ms\_ss64\_raw\_w700  & (25200, 2030, 256) & raw signals           \\
linear\_raw\_w780  & (25200, 2030, 256) & raw signals                       &  & ms\_ss64\_raw\_w730  & (25200, 2030, 256) & raw signals           \\
linear\_raw\_w800  & (25200, 2030, 256) & raw signals                       &  & ms\_ss64\_raw\_w760  & (25200, 2030, 256) & raw signals           \\
linear\_raw\_w850  & (25200, 2030, 256) & raw signals                       &  & ms\_ss64\_raw\_w730  & (25200, 2030, 256) & raw signals           \\
ms\_raw\_w700      & (25200, 2030, 256) & raw signals                       &  & ms\_ss64\_raw\_w760  & (25200, 2030, 256) & raw signals           \\
ms\_raw\_w730      & (25200, 2030, 256) & raw signals                       &  & ms\_ss64\_raw\_w780  & (25200, 2030, 256) & raw signals           \\
ms\_raw\_w760      & (25200, 2030, 256) & raw signals                       &  & ms\_ss64\_raw\_w800  & (25200, 2030, 256) & raw signals           \\
ms\_raw\_w780      & (25200, 2030, 256) & raw signals                       &  & ms\_ss64\_raw\_w850  & (25200, 2030, 256) & raw signals           \\
ms\_raw\_w800      & (25200, 2030, 256) & raw signals                       &  & ms\_ss128\_raw\_w700 & (25200, 2030, 256) & raw signals           \\
ms\_raw\_w850      & (25200, 2030, 256) & raw signals                       &  & ms\_ss128\_raw\_w730 & (25200, 2030, 256) & raw signals           \\
ms\_ss32\_BP\_w700 & (25200, 256, 256)  & image reconstructions             &  & ms\_ss128\_raw\_w760 & (25200, 2030, 256) & raw signals           \\
ms\_ss32\_BP\_w730 & (25200, 256, 256)  & image reconstructions             &  & ms\_ss128\_raw\_w780 & (25200, 2030, 256) & raw signals           \\
ms\_ss32\_BP\_w760 & (25200, 256, 256)  & image reconstructions             &  & ms\_ss128\_raw\_w800 & (25200, 2030, 256) & raw signals           \\
ms\_ss32\_BP\_w780 & (25200, 256, 256)  & image reconstructions             &  & ms\_ss128\_raw\_w850 & (25200, 2030, 256) & raw signals           \\
ms\_ss32\_BP\_w800 & (25200, 256, 256)  & image reconstructions             &  &                      &                    &                       \\
ms\_ss32\_BP\_w850 & (25200, 256, 256)  & image reconstructions             &  &                      &                    &                       \\ \bottomrule
\end{tabular}
\label{tab:msfd_contents}
\end{table}

\subsection{SWFD}

Contents of SWFD are split into two separate HDF5 files with contents listed in Tables~\ref{tab:swfd_sc_contents}~\&~\ref{tab:swfd_ms_contents} (``SWFD\_semicircle\_RawBP.h5'' and ``SWFD\_multisegment\_RawBP.h5'').
Despite having the same number of instances per dataset category across the two files, the indices are not paired across semi circle and multisegment array. 
Hence we decided to split the dataset into two files.
Furthermore, we include two additional HDF5 files ``SWFD\_semicircle\_ss\_RawBP.h5'' and ``SWFD\_multisegment\_ss\_RawBP.h5'' that contain sparse sampling image and raw data for multisegment array.
As an artifact from discarding first few slices to filter out low SNR instances (both in MSFD and SWFD), SWFD acquisitions have 1,401 slices, having \textit{sliceID} in the range $[100, ..., 1500]$, yielding $2 \times 14 \times 1$,$401 = 39$,$228$ samples per dataset category as opposed to 39,200.

\begin{table}[]
\caption{Contents of SWFD semi circle file.}
\tiny
\centering
\begin{tabular}{@{}lllllll@{}}
\toprule
Name         & Shape             & Content                              &  & Name          & Shape              & Content     \\ \cmidrule(r){1-3} \cmidrule(l){5-7} 
patientID    & (39228,)          & $(1,2,3,4,5,6,7,8,9,10,11,12,13,14)$ &  & sc,lv128\_raw & (39228, 2030, 256) & raw signals \\
side         & (39228,)          & $(\mathrm{left}, \mathrm{right})$    &  & sc,ss128\_raw & (39228, 2030, 256) & raw signals \\
skin\_type   & (39228,)          & $(1,2,3,4)$                          &  & sc,ss32\_raw  & (39228, 2030, 256) & raw signals \\
sliceID      & (39228,)          & $(100,...,1500)$                     &  & sc,ss64\_raw  & (39228, 2030, 256) & raw signals \\
sc,lv128\_BP & (39228, 256, 256) & image reconstructions                &  & sc\_raw       & (39228, 2030, 256) & raw signals \\
sc,ss128\_BP & (39228, 256, 256) & image reconstructions                &  &               &                    &             \\
sc,ss32\_BP  & (39228, 256, 256) & image reconstructions                &  &               &                    &             \\
sc,ss64\_BP  & (39228, 256, 256) & image reconstructions                &  &               &                    &             \\
sc\_BP       & (39228, 256, 256) & image reconstructions                &  &               &                    &             \\ \bottomrule
\end{tabular}
\label{tab:swfd_sc_contents}
\end{table}

\begin{table}[]
\caption{Contents of SWFD multisegment file.}
\tiny
\centering
\begin{tabular}{@{}lllllll@{}}
\toprule
Name        & Shape              & Content                              &  & Name          & Shape              & Content               \\ \cmidrule(r){1-3} \cmidrule(l){5-7} 
patientID   & (39228,)           & $(1,2,3,4,5,6,7,8,9,10,11,12,13,14)$ &  & ms,ss128\_BP  & (39228, 256, 256)  & image reconstructions \\
side        & (39228,)           & $(\mathrm{left}, \mathrm{right})$    &  & ms,ss32\_BP   & (39228, 256, 256)  & image reconstructions \\
skin\_type  & (39228,)           & $(1,2,3,4)$                          &  & ms,ss64\_BP   & (39228, 256, 256)  & image reconstructions \\
sliceID     & (39228,)           & $(100,...,1500)$                     &  & ms,ss128\_raw & (39228, 2030, 256) & raw signals           \\
linear\_BP  & (39228, 256, 256)  & image reconstructions                &  & ms,ss32\_raw  & (39228, 2030, 256) & raw signals           \\
ms\_BP      & (39228, 256, 256)  & image reconstructions                &  & ms,ss64\_raw  & (39228, 2030, 256) & raw signals           \\
linear\_raw & (39228, 2030, 256) & raw signals                          &  &               &                    &                       \\
ms\_raw     & (39228, 2030, 256) & raw signals                          &  &               &                    &                       \\ \bottomrule
\end{tabular}
\label{tab:swfd_ms_contents}
\end{table}

\subsection{SCD}

Contents of SCD are listed in Table~\ref{tab:scd_contents} and stored in the file ``SCD\_RawBP.h5''.
Different than the experimental datasets, SCD also contains the ground truth acoustic pressure map and annotations (labels).
Annotations are encoded as $0$:~background, $1$:~vessel, and $2$:~skin curve.

\begin{table}[]
\caption{Contents of SCD files.}
\tiny
\centering
\begin{tabular}{@{}lllllll@{}}
\toprule
Name          & Shape              & Content               &  & Name          & Shape               & Content               \\ \cmidrule(r){1-3} \cmidrule(l){5-7} 
sliceID       & (20000,)           & $(0 ,..., 19999)$     &  & vc,lv128\_raw & (20000, 2030, 1024) & raw signals           \\
ground\_truth & (20000, 256, 256)  & acoustic pressure map &  & vc,ss128\_raw & (20000, 2030, 1024) & raw signals           \\
labels        & (20000, 256, 256)  & $(0, 1, 2)$           &  & vc,ss32\_raw  & (20000, 2030, 1024) & raw signals           \\
linear\_BP    & (20000, 256, 256)  & image reconstructions &  & vc,ss64\_raw  & (20000, 2030, 1024) & raw signals           \\
ms\_BP        & (20000, 256, 256)  & image reconstructions &  & vc\_raw       & (20000, 2030, 1024) & raw signals           \\
linear\_raw   & (20000, 2030, 256) & raw signals           &  & ms,ss128\_BP  & (20000, 256, 256)   & image reconstructions \\
ms\_raw       & (20000, 2030, 256) & raw signals           &  & ms,ss32\_BP   & (20000, 256, 256)   & image reconstructions \\
vc,lv128\_BP  & (20000, 256, 256)  & image reconstructions &  & ms,ss64\_BP   & (20000, 256, 256)   & image reconstructions \\
vc,ss128\_BP  & (20000, 256, 256)  & image reconstructions &  & ms,ss128\_raw & (20000, 2030, 256)  & raw signals           \\
vc,ss32\_BP   & (20000, 256, 256)  & image reconstructions &  & ms,ss32\_raw  & (20000, 2030, 256)  & raw signals           \\
vc,ss64\_BP   & (20000, 256, 256)  & image reconstructions &  & ms,ss64\_raw  & (20000, 2030, 256)  & raw signals           \\
vc\_BP        & (20000, 256, 256)  & image reconstructions &  &               &                     &                       \\ \bottomrule
\end{tabular}
\label{tab:scd_contents}
\end{table}

\subsection{OADAT-mini}

OADAT-mini consists of a subset of 100 instances of each category of OADAT along with their annotation maps for vessels. 
Accordingly, its contents are split into multiple files, one for each transducer array and SCD: ``SWFD\_semicircle\_RawBP-mini.h5'', ``SWFD\_multisegment\_RawBP-mini.h5'', ``MSFD\_multisegment\_RawBP-mini.h5'', ``SCD\_RawBP-mini.h5''. 
We list the contents of each of these files in Table~\ref{tab:oadatmini_contents}.

\begin{table}[]
\caption{Contents of OADAT-mini files.}
\tiny
\centering
\begin{tabular}{@{}lllllll@{}}
\toprule
Name          & Shape              & Content               &  & Name          & Shape               & Content               \\ \cmidrule(r){1-3} \cmidrule(l){5-7} 
\quad\textbf{SCD - mini} &                    &                       &  &               &                     &                       \\
sliceID       & (100,)           & $\subset (0 ,..., 19999)$     &  & vc,lv128\_raw & (100, 2030, 1024) & raw signals           \\
ground\_truth & (100, 256, 256)  & acoustic pressure map &  & vc,ss128\_raw & (100, 2030, 1024) & raw signals           \\
labels        & (100, 256, 256)  & $(0, 1, 2)$           &  & vc,ss32\_raw  & (100, 2030, 1024) & raw signals           \\
linear\_BP    & (100, 256, 256)  & image reconstructions &  & vc,ss64\_raw  & (100, 2030, 1024) & raw signals           \\
ms\_BP        & (100, 256, 256)  & image reconstructions &  & vc\_raw       & (100, 2030, 1024) & raw signals           \\
linear\_raw   & (100, 2030, 256) & raw signals           &  & ms,ss128\_BP  & (100, 256, 256)   & image reconstructions \\
ms\_raw       & (100, 2030, 256) & raw signals           &  & ms,ss32\_BP   & (100, 256, 256)   & image reconstructions \\
vc,lv128\_BP  & (100, 256, 256)  & image reconstructions &  & ms,ss64\_BP   & (100, 256, 256)   & image reconstructions \\
vc,ss128\_BP  & (100, 256, 256)  & image reconstructions &  & ms,ss128\_raw & (100, 2030, 256)  & raw signals           \\
vc,ss32\_BP   & (100, 256, 256)  & image reconstructions &  & ms,ss32\_raw  & (100, 2030, 256)  & raw signals           \\
vc,ss64\_BP   & (100, 256, 256)  & image reconstructions &  & ms,ss64\_raw  & (100, 2030, 256)  & raw signals           \\
vc\_BP        & (100, 256, 256)  & image reconstructions &  &               &                     &                       \\ \midrule

\quad\textbf{SWFD semi circle - mini} &                    &                       &  &               &                     &                       \\
patientID    & (100,)          & $(1,2,3,4,5,6,7,8,10,11,12,13)$ &  & sc,lv128\_raw & (100, 2030, 256) & raw signals \\
side         & (100,)          & $(\mathrm{left}, \mathrm{right})$    &  & sc,ss128\_raw & (100, 2030, 256) & raw signals \\
skin\_type   & (100,)          & $(1,2,3,4)$                          &  & sc,ss32\_raw  & (100, 2030, 256) & raw signals \\
sliceID      & (100,)          & $\subset (100,...,1500)$                     &  & sc,ss64\_raw  & (100, 2030, 256) & raw signals \\
labels        & (100, 256, 256)  & $(0, 1)$ & & & & \\
sc,lv128\_BP & (100, 256, 256) & image reconstructions                &  & sc\_raw       & (100, 2030, 256) & raw signals \\
sc,ss128\_BP & (100, 256, 256) & image reconstructions                &  &               &                    &             \\
sc,ss32\_BP  & (100, 256, 256) & image reconstructions                &  &               &                    &             \\
sc,ss64\_BP  & (100, 256, 256) & image reconstructions                &  &               &                    &             \\
sc\_BP       & (100, 256, 256) & image reconstructions                &  &               &                    &             \\\midrule

\quad\textbf{SWFD multisegment - mini} &                    &                       &  &               &                     &                       \\
patientID   & (100,)           & $(1,2,3,4,5,6,7,8,9,10,11,12)$ &  & ms,ss128\_BP  & (100, 256, 256)  & image reconstructions \\
side        & (100,)           & $(\mathrm{left}, \mathrm{right})$    &  & ms,ss32\_BP   & (100, 256, 256)  & image reconstructions \\
skin\_type  & (100,)           & $(1,2,3,4)$                          &  & ms,ss64\_BP   & (100, 256, 256)  & image reconstructions \\
sliceID     & (100,)           & $\subset (100,...,1500)$                     &  & ms,ss128\_raw & (100, 2030, 256) & raw signals           \\
labels        & (100, 256, 256)  & $(0, 1)$ & & & & \\
linear\_BP  & (100, 256, 256)  & image reconstructions                &  & ms,ss32\_raw  & (100, 2030, 256) & raw signals           \\
ms\_BP      & (100, 256, 256)  & image reconstructions                &  & ms,ss64\_raw  & (100, 2030, 256) & raw signals           \\
linear\_raw & (100, 2030, 256) & raw signals                          &  &               &                    &                       \\
ms\_raw     & (100, 2030, 256) & raw signals                          &  &               &                    &                       \\ \midrule

\quad\textbf{MSFD multisegment - mini} &                    &                       &  &               &                     &                       \\
patientID          & (100,)           & $(2,5,6,7,9,10,11,14,15)$         &  & ms\_ss64\_BP\_w700   & (100, 256, 256)  & image reconstructions \\
side               & (100,)           & $(\mathrm{left}, \mathrm{right})$ &  & ms\_ss64\_BP\_w730   & (100, 256, 256)  & image reconstructions \\
skin\_type         & (100,)           & $(1,2,3,4)$                       &  & ms\_ss64\_BP\_w760   & (100, 256, 256)  & image reconstructions \\
sliceID            & (100,)           & $\subset (1,...,1400)$                    &  & ms\_ss64\_BP\_w780   & (100, 256, 256)  & image reconstructions \\
labels        & (100, 256, 256)  & $(0, 1)$ & & & & \\
linear\_BP\_w700   & (100, 256, 256)  & image reconstructions             &  & ms\_ss64\_BP\_w800   & (100, 256, 256)  & image reconstructions \\
linear\_BP\_w730   & (100, 256, 256)  & image reconstructions             &  & ms\_ss64\_BP\_w850   & (100, 256, 256)  & image reconstructions \\
linear\_BP\_w760   & (100, 256, 256)  & image reconstructions             &  & ms\_ss128\_BP\_w700  & (100, 256, 256)  & image reconstructions \\
linear\_BP\_w780   & (100, 256, 256)  & image reconstructions             &  & ms\_ss128\_BP\_w730  & (100, 256, 256)  & image reconstructions \\
linear\_BP\_w800   & (100, 256, 256)  & image reconstructions             &  & ms\_ss128\_BP\_w760  & (100, 256, 256)  & image reconstructions \\
linear\_BP\_w850   & (100, 256, 256)  & image reconstructions             &  & ms\_ss128\_BP\_w780  & (100, 256, 256)  & image reconstructions \\
ms\_BP\_w700       & (100, 256, 256)  & image reconstructions             &  & ms\_ss128\_BP\_w800  & (100, 256, 256)  & image reconstructions \\
ms\_BP\_w730       & (100, 256, 256)  & image reconstructions             &  & ms\_ss128\_BP\_w850  & (100, 256, 256)  & image reconstructions \\
ms\_BP\_w760       & (100, 256, 256)  & image reconstructions             &  & ms\_ss32\_raw\_w700  & (100, 2030, 256) & raw signals           \\
ms\_BP\_w780       & (100, 256, 256)  & image reconstructions             &  & ms\_ss32\_raw\_w730  & (100, 2030, 256) & raw signals           \\
ms\_BP\_w800       & (100, 256, 256)  & image reconstructions             &  & ms\_ss32\_raw\_w760  & (100, 2030, 256) & raw signals           \\
ms\_BP\_w850       & (100, 256, 256)  & image reconstructions             &  & ms\_ss32\_raw\_w780  & (100, 2030, 256) & raw signals           \\
linear\_raw\_w700  & (100, 2030, 256) & raw signals                       &  & ms\_ss32\_raw\_w800  & (100, 2030, 256) & raw signals           \\
linear\_raw\_w730  & (100, 2030, 256) & raw signals                       &  & ms\_ss32\_raw\_w850  & (100, 2030, 256) & raw signals           \\
linear\_raw\_w760  & (100, 2030, 256) & raw signals                       &  & ms\_ss64\_raw\_w700  & (100, 2030, 256) & raw signals           \\
linear\_raw\_w780  & (100, 2030, 256) & raw signals                       &  & ms\_ss64\_raw\_w730  & (100, 2030, 256) & raw signals           \\
linear\_raw\_w800  & (100, 2030, 256) & raw signals                       &  & ms\_ss64\_raw\_w760  & (100, 2030, 256) & raw signals           \\
linear\_raw\_w850  & (100, 2030, 256) & raw signals                       &  & ms\_ss64\_raw\_w730  & (100, 2030, 256) & raw signals           \\
ms\_raw\_w700      & (100, 2030, 256) & raw signals                       &  & ms\_ss64\_raw\_w760  & (100, 2030, 256) & raw signals           \\
ms\_raw\_w730      & (100, 2030, 256) & raw signals                       &  & ms\_ss64\_raw\_w780  & (100, 2030, 256) & raw signals           \\
ms\_raw\_w760      & (100, 2030, 256) & raw signals                       &  & ms\_ss64\_raw\_w800  & (100, 2030, 256) & raw signals           \\
ms\_raw\_w780      & (100, 2030, 256) & raw signals                       &  & ms\_ss64\_raw\_w850  & (100, 2030, 256) & raw signals           \\
ms\_raw\_w800      & (100, 2030, 256) & raw signals                       &  & ms\_ss128\_raw\_w700 & (100, 2030, 256) & raw signals           \\
ms\_raw\_w850      & (100, 2030, 256) & raw signals                       &  & ms\_ss128\_raw\_w730 & (100, 2030, 256) & raw signals           \\
ms\_ss32\_BP\_w700 & (100, 256, 256)  & image reconstructions             &  & ms\_ss128\_raw\_w760 & (100, 2030, 256) & raw signals           \\
ms\_ss32\_BP\_w730 & (100, 256, 256)  & image reconstructions             &  & ms\_ss128\_raw\_w780 & (100, 2030, 256) & raw signals           \\
ms\_ss32\_BP\_w760 & (100, 256, 256)  & image reconstructions             &  & ms\_ss128\_raw\_w800 & (100, 2030, 256) & raw signals           \\
ms\_ss32\_BP\_w780 & (100, 256, 256)  & image reconstructions             &  & ms\_ss128\_raw\_w850 & (100, 2030, 256) & raw signals           \\
ms\_ss32\_BP\_w800 & (100, 256, 256)  & image reconstructions             &  &                      &                    &                       \\
ms\_ss32\_BP\_w850 & (100, 256, 256)  & image reconstructions             &  &                      &                    &                       \\ 
\bottomrule
\end{tabular}
\label{tab:oadatmini_contents}
\end{table}

\newpage
\section{Considerations when using our datasets}

As the volunteers in experimental datasets are considered to be healthy, anonymization is done one way, and true identities of volunteers are not possible to trace. 
Furthermore, to the best of our knowledge, usage of our proposed datasets cannot pose threat to the volunteers, even with malicious intent. 
However, as with all clinical data, it should be acknowledged that our dataset would represent a particular subset of all potential forearm images collected at respective wavelengths and transducer arrays. 
Accordingly, any subsequent work that makes use of our datasets for validation purposes need to ensure that the diversity of our datasets is sufficient for their target application.

\section{Persistence of proposed datasets}

All datasets are hosted on Libdrive at ETH Zurich Research Collection repository.
Accordingly, the datasets have a landing page with DOI and metadata. 
Data will be freely accessible with no restriction. 
As per ETH Zurich Research Collection documentation, the dataset is guaranteed to have a retention period for 10 years. 
Based on the frequency of usage by the community, the Research Collection may continue to host the data beyond the 10 year period.

\section{Author statement on data license}

The dataset is licensed under Creative Commons Attribution-NonCommercial 4.0 International (CC-BY-NC).

\section{Training and evaluation code}

In \href{https://renkulab.io/gitlab/firat.ozdemir/oadat-evaluate}{oadat-evaluate repository}, along with saved model weights, we provide all necessary scripts to train modUNet from scratch for all 44 experiments as well as evaluating them over the whole test set. 
We provide standalone examples on the repository landing page to show how to;
\\
(i) load a pretrained model, for example, to do inference or finetuning it,  
\\
(ii) train modUNet from scratch for either of the two tasks for a given experiment,
\\
(iii) evaluate a serialized model, whether it is one of the pretrained models we provide as is, or any other Tensorflow model that is already serialized, and
\\
(iv) use our provided data loader to read any sample from OADAT.

\end{document}